\documentclass[prl,twocolumn,noshowpacs,preprintnumbers,amsmath,amssymb,nofootinbib,showkeys]{revtex4-2}
\usepackage{bbm}
\usepackage{lipsum}
\usepackage{mathrsfs}
\usepackage{graphicx}
\usepackage{dcolumn}
\usepackage{bm}
\usepackage{color}
\usepackage{comment}
\usepackage{hyperref}
\usepackage{times}
\usepackage{ulem}
\usepackage{footnote}

\newcommand{\tb}{\textbf}

\newcommand{\Day}{\rm D}

\newcommand{\ra}{\rangle}
\newcommand{\la}{\langle}
\newcommand{\dla}{\la\!\la}
\newcommand{\dra}{\ra\!\ra}

\newcommand{\eps}{\epsilon}

\newcommand{\mcQ}{\mathcal{Q}}
\newcommand{\mcI}{\mathcal{I}}

\newcommand{\SM}{Supplemental Material}
\newcommand{\SMabb}{SM}

\begin{document}
\title{
		Strict universality of the square-root law in price impact across stocks:\\
		a complete survey of the Tokyo stock exchange
}
\author{Yuki Sato}

\author{Kiyoshi Kanazawa$^*$}

\affiliation{Department of Physics, Graduate School of Science, Kyoto University, Kyoto 606-8502, Japan}
\date{\today}
\begin{abstract}
	Universal power laws have been scrutinised in physics and beyond, and a long-standing debate exists in econophysics regarding the strict universality of the nonlinear price impact, commonly referred to as the square-root law (SRL). The SRL posits that the average price impact $I$ follows a power law with respect to transaction volume $Q$, such that $I(Q) \propto Q^{\delta}$ with $\delta \approx 1/2$. Some researchers argue that the exponent $\delta$ should be system-specific, without universality. Conversely, others contend that $\delta$ should be exactly $1/2$ for all stocks across all countries, implying universality. However, resolving this debate requires high-precision measurements of $\delta$ with errors of around $0.1$ across hundreds of stocks, which has been extremely challenging due to the scarcity of large microscopic datasets---those that enable tracking the trading behaviour of all individual accounts. Here we conclusively support the universality hypothesis of the SRL by a complete survey of all trading accounts for all liquid stocks on the Tokyo Stock Exchange (TSE) over eight years. Using this comprehensive microscopic dataset, we show that the exponent $\delta$ is equal to $1/2$ within statistical errors at both the individual stock level and the individual trader level. Additionally, we rejected two prominent models supporting the nonuniversality hypothesis: the Gabaix-Gopikrishnan-Plerou-Stanley and the Farmer-Gerig-Lillo-Waelbroeck models (Nature 2003, QJE 2006, and Quant. Finance 2013). Our work provides exceptionally high-precision evidence for the universality hypothesis in social science and could prove useful in evaluating the price impact by large investors---an important topic even among practitioners.	
\end{abstract}


\maketitle

	\paragraph*{Introduction.}

	\begin{figure*}
		\includegraphics[width=165mm]{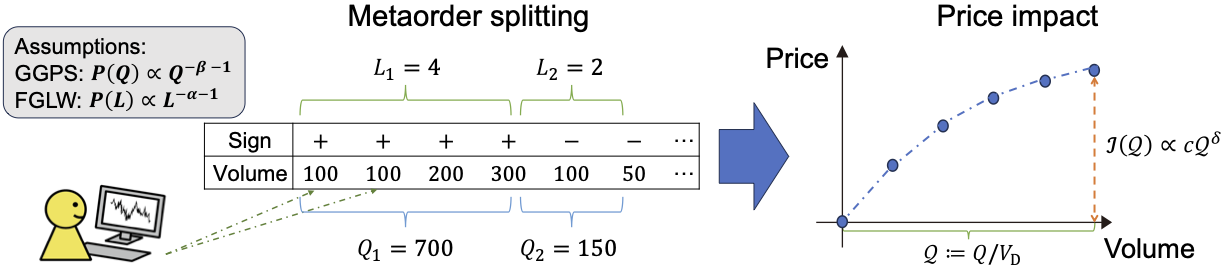}
		\caption{
			Square-root price impact law.
			Practitioners typically split a large volume of market orders (called a metaorder, with size $Q$) into a long sequence of child orders (with length $L$).
			Price impact $I(Q):= \la \eps \Delta p \mid Q\ra$ is the average mid-price change $\Delta p$ between the beginning and the end of a buy (sell) metaorder with $\eps=+1$ ($\eps=-1$) and follows $I(Q)\propto Q^{\delta}$ with $\delta\approx 1/2$.
		}
		\label{fig:Schematic}
	\end{figure*}

	Universality is one of the central topics in physics, whereby common macroscopic laws are observed across various materials~\cite{StanleyB1971,GoldenfeldB}. For example, universal power laws are observed in critical phenomena, such as $M \propto |T - T_c|^{\delta}$, with an order parameter $M$ near the critical point $T_c$ (e.g., the magnetisation $M$ and temperature $T$ for a ferromagnetic system). Interestingly, the exponent $\delta$ is independent of the microscopic details of the system. Such a robust exponent is referred to as a {\it universal exponent} to emphasise the special significance of universality, in contrast to a nonuniversal exponent that may be a fragile observable, as it could easily vary with perturbations to the system's microscopic parameters. 

	Inspired by such success, physicists have attempted to discover universal power laws beyond physics, including in social science. For instance, in finance, some physicists believe the {\it square-root law} (SRL) for price impact~\cite{BouchaudText} should be part of such universal scaling laws~\cite{Toth2011,MastromatteoPRL2014,MastromatteoPRE2014,Bucci2019PRL,DonierLOB2015}. At the same time, this universality claim has not been fully established for lack of large financial microscopic datasets that enable the tracking of all trading accounts individually. Here we quantitatively support the universality hypothesis through a complete survey of all trading accounts on the Tokyo Stock Exchange (TSE), aiming to conclude the long-standing debate in econophysics. 
	
	The SRL is an empirical law that describes the nonlinear price response following large buying (selling) by institutional investors. Consider a trader buying (or selling) $Q$ units of stock by market orders---immediate decisions to buy or sell shares. Following this transaction, the market responds with a price movement $\Delta p$. For small $Q$, a linear response is expected, such that $I(Q):= \la \eps \Delta p \mid Q\ra \approx \lambda Q$, where $I(Q)$ is called the {\it price impact}~\cite{BouchaudText}, $\eps$ is the market order sign defined by $\eps=+1$ ($\eps=-1$) for buying (selling), and $\la A \mid B\ra$ represents the average of $A$ conditioned on $B$. For large $Q$, on the contrary, a nonlinear response is empirically observed (see Fig.~\ref{fig:Schematic}),
	\begin{equation}
		\mcI(\mcQ) \approx c_0 \mcQ^{\delta}\label{eq:SRL-scaling}
	\end{equation}
	after the nondimensionalisation $\mcQ:=Q/V_{\Day}$ and $\mcI(\mcQ):= I(\mcQ)/\sigma_{\Day}$, where $V_{\Day}$ is the daily transaction volume and $\sigma_{\Day}$ is the daily volatility. Surprisingly, the exponent $\delta$ is typically claimed to be close to one-half for any asset worldwide, 
	\begin{equation}
		\delta \approx \frac{1}{2}.\label{eq:SRL-exponent}
	\end{equation}
	Thus, the nonlinear scaling~\eqref{eq:SRL-scaling} is referred to as the SRL~\cite{Toth2011, MastromatteoPRL2014, MastromatteoPRE2014, Bucci2019PRL,DonierLOB2015}. For example, the SRL predicts buying 1\% of a stock's daily traded shares typically moves the price by about 10\% of its daily price range. The SRL closely resembles the universal power law observed in physics and has been an exceptionally appealing topic in econophysics (see {\SM} (\SMabb)~\cite{SM} for a detailed literature review).

	However, there has been a long-standing debate regarding the strict universality of the SRL. Some physicists argue that the exponent $\delta$ should depend on the microscopic details of financial markets and, therefore, $\delta$ should be considered a nonuniversal exponent. For instance, the Gabaix-Gopikrishnan-Plerou-Stanley (GGPS) and the Farmer-Gerig-Lillo-Waelbroeck (FGLW) models are two promising microscopic frameworks that support the nonuniversality hypothesis~\cite{NatureGabaix,Gabaix2006,FGLWQFin2013}. Conversely, other physicists contend that $\delta$ should be exactly one-half for all financial markets in any country, making $\delta$ a universal exponent. For example, the latent order-book model~\cite{Toth2011,MastromatteoPRL2014,MastromatteoPRE2014,DonierLOB2015} proposes a minimal mechanism to derive $\delta=1/2$, supporting the universality hypothesis. Which hypothesis is ultimately correct? This is a major point even in financial economics, as hypothesis selection is strongly relevant for validating/rejecting microscopic models to explain the SRL's mechanisms.

	Despite its evident importance, this debate has not yet been settled for several reasons. First, microscopic financial datasets---those that enable tracking trading behaviour for all individual trading accounts---are essential to test these hypotheses. Unfortunately, such microscopic datasets are extremely rare, making it challenging to rigorously test these hypotheses. Second, while previous studies have investigated the SRL, their datasets were typically proprietary, provided by specific hedge funds. This means that their datasets are neither comprehensive nor randomly sampled, which introduces the possibility of sampling bias, and the resulting sample sizes are inevitably small. A complete survey---studying all trading accounts for all liquid stocks---has been highly desired to address this issue convincingly. Third, solving these issues requires extremely high accuracy in measuring the exponent $\delta$. To test these hypotheses, we need to measure $\delta$ with an accuracy of about $0.1$ for hundreds of stocks. Such a high-precision measurement is difficult to achieve in social sciences for lack of large microscopic datasets.

	In this article, we present conclusive evidence supporting the universality of the SRL by a complete survey of the TSE. We analysed a large, high-quality microscopic dataset that records the percise trading behaviour of all trading accounts on the TSE for all stocks over eight years. Our TSE dataset includes virtual server identifiers (IDs)---a unit of trading accounts on the TSE---enabling us to investigate the resulting price impact $\Delta p$ associated with the buy (sell) volume $Q$ at the individual account level. Using this dataset, we precisely measured $\delta$ by controlling errors of the order of $0.06$ and found that $\delta$ was equal to $1/2$ for all liquid stocks (more than two thousand data points) within the statistical errors. We also measured $\delta$ for all active traders individually and found it to be equal to $1/2$ within statistical errors as well. Furthermore, we rejected two prominent models that had theoretically supported the nonuniversality hypothesis: the GGPS and the FGLW models. Our result supports the universality hypothesis of the SRL with extraordinary accuracy.

\paragraph*{Data.}
	Our dataset was provided by Japan Exchange (JPX) Group, Inc., the platform manager of the TSE market. This dataset records the complete life cycle of all order submissions for all stocks on the TSE over almost eight years. The unique feature of our dataset is that it includes virtual server IDs. Virtual server IDs are not technically equivalent to the membership IDs, because any trader may possess several virtual servers. However, we can define effective trader IDs, referred to as {\it trading desks}, by tracking all the virtual server IDs (see End Matter, the SM~\cite{SM}, and previous studies~\cite{Goshima2019,SatoPRL2023,SatoPRR2023}). In this article, we refer to these trading desks as trader IDs for short.

	\paragraph*{Stock-level histogram of $\delta$.} 
	\begin{figure*}
		\includegraphics[width=170mm]{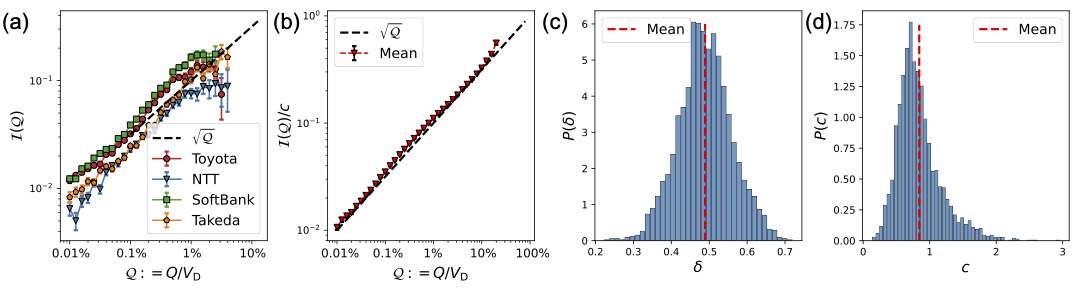}
		\caption{
			Stock-level price impact. 
			(a)~Average price-impact profile for four stocks (red: Toyota Motor Corp., blue: Nippon Telegraph and Telephone (NTT) Corp., green: SoftBank Group Corp., orange: Takeda Pharmaceutical Company Limited). Errorbars represent the standard errors of the means (SEMs).
			(b)~Aggregate scaling plot, visualising the average scaling profiles across all stocks, confirms the square-root law for all stocks.
			(c)~Histogram of the exponent $\delta$ across all stocks, exhibiting a sharp peak around $0.5$ with $\la\delta\ra=0.489$ with its SEM of $0.0015$ and $\sigma_\delta:= \sqrt{\la (\delta- \la\delta\ra)^2\ra}=0.071$. 
			(d)~Histogram of the prefactor $c$ across all stocks, with $\la c\ra=0.842$. 
			}\label{fig:DeltaStatistics}
	\end{figure*}
	We measured the transaction volume $Q$ and corresponding price impact $\Delta p$ for all liquid stocks on the TSE. First, we defined trader IDs for each stock and extracted the market-order volume sequences. Second, we assume that institutional investors typically split their large orders (called metaorders) into small orders (called child orders), as suggested by the literature~\cite{LMF_PRE2005,SatoPRL2023,SatoPRR2023,Bershova,GeneralizedLMF}. Specifically, if two successive market orders have the same sign (plus for buy and minus for sell), we assume that both orders are child orders belonging to the same metaorder. The total volume $Q$ is then defined as the sum of all child-order volumes. Third, the price impact $\Delta p$ is defined as the price movement caused by such a metaorder. In this article, liquid stocks are defined as stocks where the total number of metaorders exceeds $10^5$.

	Figure~\ref{fig:DeltaStatistics}(a) presents the typical price-impact profile $I(Q)$ for four liquid stocks. Additionally, we present Fig.~\ref{fig:DeltaStatistics}(b) as the aggregate scaling plot---visualising the averaged impact profile across all stocks as one figure.  The aggregate scaling plot was created by two steps (see End Matter and the SM~\cite{SM}): (i)~After non-dimensionalising $\mcQ:=Q/V_{\Day}$ and $\mcI(\mcQ):=I(\mcQ)/\sigma_{\Day}$ for each stock, we applied fitting $\mcI(\mcQ)=c\mcQ^{1/2}$ to estimate $c$. (ii)~The average of $\mcI(\mcQ)/c$ across all stocks was plotted, which collapsed onto a single master curve as theoretically expected. These two figures provide an overall picture of the SRL~\eqref{eq:SRL-scaling} and qualitatively support its validity across all stocks.

	We next present the normalised histogram of $\delta$ across all stocks as Fig.~\ref{fig:DeltaStatistics}(c). We estimated $\delta$ and $c$ using the power-law scaling~\eqref{eq:SRL-scaling} by applying the nonlinear relative least squares on binned averages of price impact (see SM~\cite{SM} for a robustness check on the estimation method). Due to the finite-sample size, the $\delta$'s errorbars for individual stocks are roughly given by $\dla \overline{\sigma_{\delta}}\dra=0.063$, estimated by a plausible statistical model as discussed in the next paragraph. Figure~\ref{fig:DeltaStatistics}(c) shows a very sharp peak around $1/2$ with the average and the standard deviation, 
	\begin{equation}
		\la \delta\ra=0.489 \pm 0.0015, \>\>\> \sigma_\delta:= \sqrt{\la (\delta- \la\delta\ra)^2\ra}= 0.071,
	\end{equation}
	where $\la A\ra$ represents the cross-sectional average of $A$ on the TSE dataset across all stocks\footnote{
		If we estimate $\delta$ by the nonlinear ordinary least squares on raw datapoints (without using binned averages), we obtain $\la \delta_{\rm OLS}\ra=0.474 \pm 0.0012$.
	}. This is the first main result of this article, suggesting that the exact SRL holds for more than two thousand data points. Also, Fig.~\ref{fig:DeltaStatistics}(d) presents the histogram of the prefactor $c$. Furthermore, the crossover from the linear to square-root laws was observed consistently with Refs.~\cite{Zarinelli2015,Bucci2019PRL} (see SM~\cite{SM}).

\paragraph{Errorbar estimation.}
	Statistically, estimating errorbars is necessary for $\delta$ due to finite sample size. If all the observations of $(\mcQ,\mcI)$ were independent and identically distributed (IID), simple nonlinear regression would suffice to estimate them. However, our dataset is a time series with serial correlation. For non-IID observations, estimating errorbars requires plausible statistical models to account for serial correlation (e.g., the autoregressive integrated moving average (ARIMA) model~\cite{JDHamiltonB}). Thus, developing an appropriate statistical model for our dataset is crucial for obtaining reliable error estimates\footnote{
		If we ignore the non-IID nature of the data and use standard error estimation packages (e.g., \texttt{scipy.optimize.least\_squares} on Python~\cite{Scipy}), the errorbar of $\delta$ is estimated to be $0.022$ via the nonlinear least relative squares on binned averages, or $0.014$ via the nonlinear ordinary least squares on raw datapoints. Both of these values are underestimates of the true errorbar in the presence of serial correlation.
	}.

	\begin{figure}
		\centering
		\includegraphics[width=85mm]{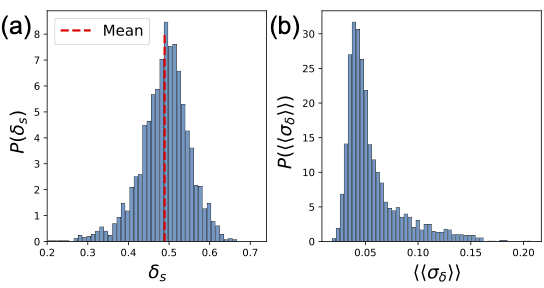}
		\caption{
			Errorbar estimation based on our statistical model. 
			(a)~$\delta$'s histgram across all stocks for one Monte Carlo iteration, consistent with the $\delta$'s histgram~\ref{fig:DeltaStatistics}(c) for our TSE dataset.
			(b)~Histgram of $\delta$'s errorbars $\dla \sigma_{\delta}\dra$ for individual stocks by the repeated Monte Carlo simulation. Each errorbar was estimated with $100$ iterations.
		}
		\label{fig:errorbar-statistical-model}
	\end{figure}
	For this purpose, we develop a simple statistical model with the exact SRL under order splitting and estimate the corresponding errorbar (see End Matter and the SM~\cite{SM}): (i)~We randomly shuffled the signs of metaorders, while keeping the physical times of market-order submissions identical to those in our TSE dataset. (ii)~The price impacts were assumed to exactly obey the SRL with $\delta=1/2$, with random contributions. The model parameters were based on our dataset. (iii)~We repeatedly generated price time series for 100 iterations and measured $\delta$ for all stocks (see Fig.~\ref{fig:errorbar-statistical-model}(a) for a $\delta$'s histogram for one numerical iteration, corresponding to Fig.~\ref{fig:DeltaStatistics}(c) for our TSE dataset). (iv)~The average and standard deviation of $\delta$ were evaluated across all Monte Carlo trials as $\dla \overline{\delta}\dra$ and $\dla \overline{\sigma_{\delta}}\dra$. For our numerical simulation, $\dla A \dra$ represents the Monte Carlo average of $A$ over $N_{\rm MC}=100$ iterations, whereas $\overline{A}$ represents the average across stocks for one Monte Carlo iteration. This model generates plausible price time-series, consistent with diffusive price. Indeed, Ref.~\cite{NLProp2025} develops a nonlinear price-impact model under order splitting, reminiscent of this statistical model, to show its consistency with various stylized facts.
	
	Since $\delta=1/2$ exactly in this statistical model, the dispersion $\dla \overline{\sigma_{\delta}}\dra$ represents the errorbar due to the finite sample size in the frequentist sense. Finally, we obtained
	\begin{equation}
		\dla \overline{\delta}\dra=0.489 \pm 0.0013, \>\>\> \dla \overline{\sigma_{\delta}}\dra=0.063.
	\end{equation}
	Hense, we estimate that our $\delta$'s errorbar is given by $\dla \overline{\sigma_{\delta}}\dra=0.063$ for the first main result in Fig.~\ref{fig:DeltaStatistics}(c).	Also, see Fig.~\ref{fig:errorbar-statistical-model}(b) for the histogram of errorbars for individual stocks. 
	
	Remarkably, the cross-sectional average $\la \delta\ra$ for the TSE dataset is equal to the Monte-Carlo average $\dla \overline{\delta}\dra$ for our statistical model within the errorbar, such that $|\la \delta\ra-\dla \overline{\delta}\dra|\ll \dla \overline{\sigma_{\delta}}\dra$. In addition, the cross-sectional dispersion $\sigma_\delta$ for the TSE dataset is almost equal to that $\dla \overline{\sigma_{\delta}}\dra$ for our statistical model, such that $\sigma_\delta\approx \dla \overline{\sigma_{\delta}}\dra$. These facts imply the self-consistency of our statistical analysis. Thus, we concluded that the standard deviation in estimating $\delta$ from our dataset is almost attributable to finite sample size, and the universality hypothesis holds for the SRL within statistical errors, at least for the TSE.

\paragraph*{Trader-level histogram of $\delta$.} 
	\begin{figure*}
		\includegraphics[width=170mm]{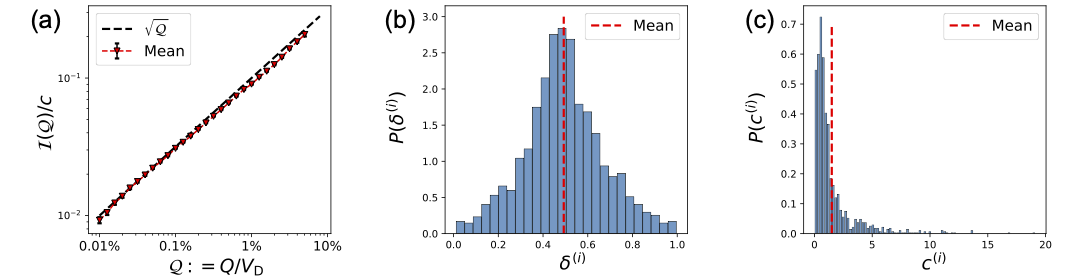}
		\caption{
			Trader-level price impact. 
			(a)~Trader-level aggregate scaling plot confirms the square-root law for all traders.
			(b)~Histogram of the exponent $\delta^{(i)}$ across all traders for $0 \leq \delta^{(i)} \leq 1$ shows a sharp peak around $0.5$, with $\la \delta^{(i)} \ra = 0.493$ with its SEM of $0.0050$ and $\sigma_{\delta^{(i)}} := \sqrt{\la(\delta^{(i)} - \la\delta^{(i)}\ra)^2\ra} = 0.177$. 
			A few outliers were observed outside this region due to fitting errors.
			(c)~Histogram of the prefactor $c^{(i)}$ across all traders, with $\la c^{(i)}\ra=1.501$. 
		}
		\label{fig:Trader}
	\end{figure*}
	While we presented evidence at the stock level, is the SRL universal even at the trader level? In other words, by writing the exponent for the $i$-th trader as $\delta^{(i)}$, is $\delta^{(i)}$ statistically indistinguishable from $1/2$ for all traders, despite the diversity in trading strategies among traders? To answer this question, we studied the price impact of all active traders---roughly defined as traders who submitted metaorders not less than $10^4$ times in this article (see End Matter and SM~\cite{SM}). With this definition, we have 1,293 active traders in total. We present the aggregate scaling plot at the individual trader level as Fig.~\ref{fig:Trader}(a), qualitatively suggesting the validity of the SRL for all traders. As a second main result, we present the normalised trader-level histogram of $\delta^{(i)}$ as Fig.~\ref{fig:Trader}(b). The histogram has clear sharp peak with the average $\la \delta^{(i)}\ra$ across all traders on the TSE dataset and standard deviation $\sigma_{\delta^{(i)}}:= \sqrt{\la (\delta^{(i)}- \la\delta^{(i)}\ra)^2\ra}$ as
	\begin{equation}
		\la \delta^{(i)}\ra=0.493\pm 0.0050, \>\>\> \sigma_{\delta^{(i)}}= 0.177,
	\end{equation}
	quantitatively supporting the SRL $\la \delta^{(i)}\ra\approx 1/2$ within statistical errors. As a reference, we plot the histogram of the prefactor $c^{(i)}$ across all traders in Fig.~\ref{fig:Trader}(c).

	The standard deviation $\sigma_{\delta^{(i)}}$ in our data was at the same level as that evaluated by our numerical model, $\dla \overline{\delta^{(i)}}\dra=0.521\pm 0.0048$ and $\dla \overline{\sigma_{\delta^{(i)}}}\dra=0.169$, where $\overline{A}$ represents the average across all traders in our numerical simulation. Since $|\la \delta^{(i)}\ra-\dla \overline{\delta^{(i)}}\dra|\ll \dla \overline{\sigma_{\delta^{(i)}}}\dra$ and $\sigma_{\delta^{(i)}}\approx \dla \overline{\sigma_{\delta^{(i)}}}\dra$, we concluded that the universality hypothesis of the SRL is maintained within statistical errors even at the trader level. 

\paragraph*{Rejecting two nonuniversal models.} 
	\begin{figure*}
		\includegraphics[width=170mm]{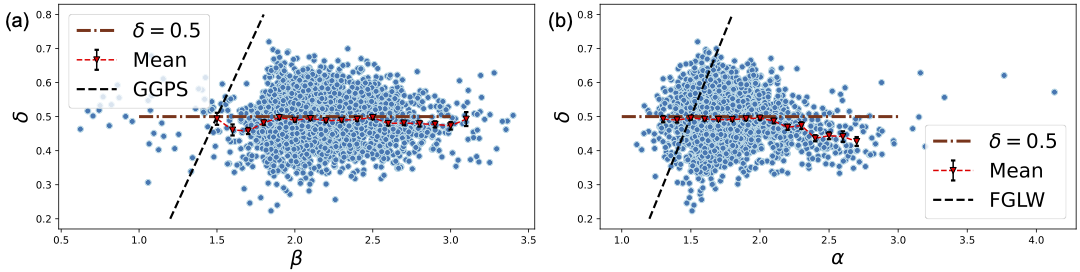}
		\caption{
			Quantitative test of (a)~the GGPS prediction $\delta=\beta-1$ and for (b)~the FGLW prediction $\delta=\alpha-1$. No correlation was found and, thus, both models were rejected.
			}
		\label{fig:FGLW-GGPW}
	\end{figure*}
	While we presented direct evidence supporting the universality hypothesis, we now present direct negative evidence against two prominent models supporting the nonuniversality hypothesis: the GGPS and the FGLW models~\cite{NatureGabaix,Gabaix2006,FGLWQFin2013}.

	The GGPS and FGLW models are based on microeconomic theory in the presence of order-splitters: Suppose that an order-splitter has a metaorder with total volume $Q$ and splits it into $L$ child orders. Empirically, the distributions of the metaorder volume $Q$ and the number of corresponding child orders $L$ are known to obey power laws \cite{Bershova,SatoPRL2023,SatoPRR2023,LMF_PRE2005,NatureGabaix,Gabaix2006}: $P(Q)\propto Q^{-\beta-1}$ for large $Q$ and $P(L)\propto L^{-\alpha-1}$ for large $L$, where $\beta$ and $\alpha$ are positive power-law exponents characterising individual stocks. Based on this key assumption, the GGPS and FGLW models consider the strategy optimisation of market makers in light of inventory risk and fair pricing, respectively. They predict that the exponent $\delta$ is given by 
	\begin{equation}
		\mbox{GGPS: }\delta = \beta - 1, \>\>\> 
		\mbox{FGLW: }\delta = \alpha - 1,
		\label{eq:prediction-nonuniversal}
	\end{equation}
	where Zipf's law is assumed for the company-size distribution as for the GGPS theory. 
	
	Interestingly, the typical values of $\beta$ and $\alpha$ are distributed around 1.5, which is superficially consistent with the SRL~\cite{SatoPRL2023,SatoPRR2023,NatureGabaix,Gabaix2006,Bershova}. However, since $\beta$ and $\alpha$ can differ from 1.5 depending on the stock, the GGPS and FGLW models have suggested the theoretical nonuniversality of the exponent $\delta$. Their prediction~\eqref{eq:prediction-nonuniversal} can be tested by precisely estimating $\beta$, $\alpha$, and $\delta$. While a small group of researchers tested this relation~\eqref{eq:prediction-nonuniversal}, both positive and negative results~\cite{MastromatteoPRE2014,NatureGabaix,Gabaix2006,Bershova} coexist, as direct precise measurement without bias is not an easy task for $\beta$, $\alpha$, and $\delta$, except in a complete survey. We address this issue definitively using our large TSE dataset. 

	To test the GGPS and FGLW models, it is sufficient to present two high-accuracy scatterplots: one between $\beta$ and $\delta$, and the other between $\alpha$ and $\delta$. We present Fig.~\ref{fig:FGLW-GGPW} as our final main result. Clearly, there is no correlation at all, rejecting both of the two prominent models that have supported the universality hypothesis, at least on the TSE.

\paragraph*{Concluding discussion.}

	This Letter provides the strongest evidence so far supporting the universality hypothesis, at least on the TSE. See also the SM~\cite{SM} for thorough statistical robustness of our findings. It would be desirable to perform complete surveys of other markets, which is left as a future study. However, we stress that a single-market survey is sufficient to falsify the universal validity of the prominent non-universal models, as shown in this work.

	Economically, the square-root law implies that market liquidity is larger against metaorder splitting than expected by linear models in traditional economics~\cite{Kyle}. Additionally, Ref.~\cite{NLProp2025} implies the SRL plays a key role in validating the efficient market hypothesis---one of the most important concepts in financial economics. Thus, our study sheds new light on market liquidity in the presence of large institutional investors.

\begin{acknowledgements}
	\paragraph*{Acknowledgements.}
		We appreciate the careful reading of this manuscript by the staff of JPX Group. Also, we thank Jean-Phillipe Bouchaud for his motivating comments on our conference presentations and this manuscript. YS was supported by JSPS KAKENHI (Grant No.~24KJ1328). KK was supported by JSPS KAKENHI (Grant Nos.~21H01560, 22H01141, and 25K01450) and the JSPS Core-to-Core Program (Grant No.~JPJSCCA20200001).
\end{acknowledgements}
	
\paragraph*{Data availability.}
	The data supporting our results were provided by Japan Exchange (JPX) Group, Inc. JPX Group is a third-party commercial company and provided their dataset through a non-disclosure agreement with Kyoto University, strictly for academic purposes. This non-disclosure agreement imposes legal restrictions on data availability, and therefore, we cannot make the data publicly accessible without approval from JPX Group. However, upon reasonable request, KK will provide all the original data and programming code, subject to explicit approval by JPX Group.

\paragraph*{Code availability.}
	Our numerical simulation code, used to estimate the finite sample-size effect on the standard deviation of the estimated exponent $\delta$, is available from the following URL: \url{https://gitlab.com/Yuki-Sato_JPN/CodeDisclosure-2024-SRL}.

\paragraph*{Ethics statement}
 	The TSE dataset does not contain personal information, as defined by the Act on the Protection of Personal Information in Japan, and no ethical issues are involved in this research. JPX Group defines market participants at the corporate level. JPX Group collected trading log data at the corporate account level by making agreements with corporate-level market participants; in other words, personal information linked to individual traders behind corporate accounts was not collected. Additionally, an ethical review is not required for social science research based on such corporate-account-level datasets at Kyoto University.

\paragraph*{Author contribution.}
YS developed the program code and conducted the statistical analysis. KK designed and managed the project. Both YS and KK contributed to writing the manuscript and approved its final content.

\paragraph*{Conflict of Interest statement.}
	We declare no financial conflict of interest. JPX Group provided the original data for this study without financial support.

\clearpage


	\appendix
	\section{End Matter}

	\begin{figure*}
		\centering
		\includegraphics[width=165mm]{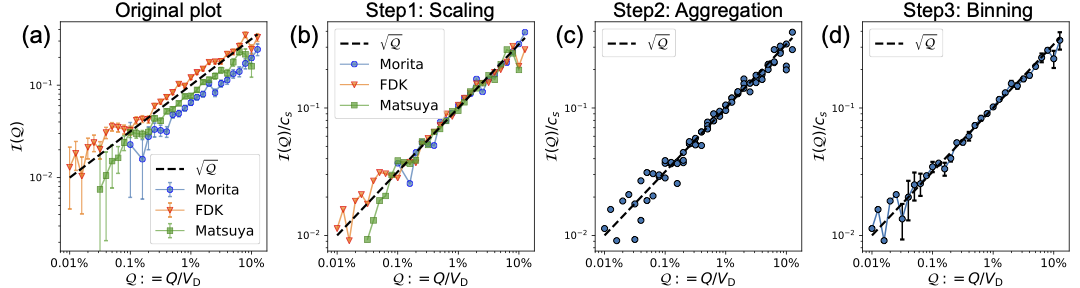}
		\caption{
			Schematic example of the aggregate scaling plots: (a)~Average price impact profile for the three specific stocks from Dataset 2: Morita Holdings Corp. (blue, ticker 6455, $s=1$), FDK Corp. (orange, ticker 6955, $s=2$), and Matsuya Co. LTD. (green, ticker 8237, $s=3$), where $s$ represents the stock index. While this data agrees with the square-root law $\mcI_s(\mcQ)=c_s\sqrt{\mcQ}$ on average, the three coefficient $\{c_s\}_{s=1,2,3}$ are not identical. (b)~This heterogeneity is removed by applying the rescaling $\phi(x)=\mcI_s(\mcQ(x))/c_{s}$ with $x:=\sqrt{\mcQ}$. (c)~Next, the average $\phi_{\rm avg}(x)$ is obtained across all the stocks. (d)~Finally, the averaged profile $\phi_{\rm avg}(x)$ is plotted with bins evenly spaced on a logarithmic scale.
			}
		\label{fig:agg_scaling_proc}
	\end{figure*}	
	\paragraph*{Appendix A: Data description.}
	We used the JPX order-book dataset on the TSE. This data covers the complete life cycle of all orders, from submissions to transactions/cancellations, for all stocks on the TSE over eight years (from 4th January 2012 to 2nd November 2019). 
		
	We divided this dataset into two datasets: one from 4 January 2012 to 18 September 2015 (Dataset 1), and another from 24 September 2015 to 2 November 2019 (Dataset 2). This is because there was a technical update on the {\it arrowhead}---the order-matching system managed by the TSE---on 24 September 2015, and all the virtual server IDs were shuffled. 

	A liquid stock is defined as one where the total number of metaorders exceeds $10^5$. This filter is introduced to control the sample size and the resulting accuracy of estimation. In addition, any stock in Dataset 1 is regarded as different from stocks in Dataset 2, even if their ticker codes are identical. With this definition, the total number of liquid stocks in Dataset 1 and Dataset 2 are 942 and 1,357, respectively. Thus, we have 2,299 datapoints in total regarding liquid stocks.

	Our TSE dataset records virtual server IDs corresponding to each order submission. A virtual server acts as an effective unit of trading accounts on the TSE, as all orders are submitted from a virtual server. Technically, a virtual server ID is not exactly technically equivalent to a membership ID, as any membership may possess several virtual servers. In fact, there is a limit on the total number of submissions that can be made during a fixed interval from one virtual server, and high-frequency traders might possess several virtual servers to circumvent this limit. However, we can allocate effective trader IDs by appropriate aggregation of virtual server IDs. 
	
	\begin{table}
		\begin{tabular}{c|c|c|c}
				\hline 
					Virtual server ID  &  Order ID  &  Order type & Effective trader ID \\
			\hline \hline
					V1  &  O1   &  Limit order  & T1 \\
					V2  &  O1   &  Cancellation & T1 \\
					V3  &  O2   &  Limit order  & T2 \\
					V3  &  O2   &  Cancellation & T2 \\
					\vdots  &  \vdots   &  \vdots  & \vdots \\
		\end{tabular}
		\caption{
			Schematic of how trader IDs are defined from virtual server IDs.
		}
	\label{method:tab:trading-desk}
	\end{table}
	Effective trader IDs are defined as follows: While the lifecycle of a single order ID is typically managed by a single virtual server, there are exceptional cases where multiple virtual servers are involved in the lifecycle of one order. Suppose that virtual server ``V1" submits a limit order, and then virtual server ``V2" cancels the same limit order under the same order ID ``O1" (see Table~\ref{method:tab:trading-desk}). We infer that the same trader operates both virtual servers, V1 and V2, and we assign a unified identifier, ``T1," as the effective trader ID. This effective trader ID is referred to as a trader ID for simplicity, while it was originally called trading desks in the literature~\cite{Goshima2019,SatoPRL2023,SatoPRR2023}.

	We define trader IDs for each stock: Even if trader ``T1'' made transactions for two different stocks (such as Toyota and NTT), we define two traders, such as ``Dataset1-Toyota-T1'' and ``Dataset1-NTT-T1.''

\paragraph*{Appendix B: Definition of metaorders.}
Metaorders are defined based on the order-splitting hypothesis: typical large investors split a large metaorder into a long sequence of child orders. For example, trader ``T1'' may have a buy (sell) metaorder with a volume of $Q=+100$ ($Q=-100$) and split it into ten buy (sell) child orders, such as $q_1=q_2=...=q_{10}=+10$ ($q_1=q_2=...=q_{10}=-10$), where the sign ``$+$" (``$-$")	represents a buy (sell) order. 
	
	Based on this hypothesis, we assume that any two successive orders belong to the same metaorder if their order signs are identical. For example, if we have a market order sequence $\{q_k^{(i)}\}_k = \{+10,+10,+1,+5,-3,-2,-1, +100\}$ for trader $i$, the $k$-th metaorder volume $Q_k^{(i)}$ and the corresponding number of child orders $L_k^{(i)}$ are given by $(Q_1^{(i)},L_1^{(i)})=(+26, 4)$, $(Q_2^{(i)},L_2^{(i)})=(-6,3)$, and $(Q_3^{(i)},L_3^{(i)})=(+100,1)$. Additionally, any metaorder is defined on a daily basis: market orders on different days are regarded as belonging to separate metaorders.

	With this definition, we determined the metaorder volumes and lengths for each stock, and then measured the power-law exponents $\beta$ and $\alpha$ using Clauset's algorithm~\cite{Clauset}. These exponents were used to draw the scatterplots in Fig.~\ref{fig:FGLW-GGPW}.

\paragraph*{Appendix C: Exceptional handling.}
	The TSE has exceptional rules known as the {\it daily price limit} and {\it circuit breaker}, which are aimed at protecting investors from sudden price jumps. Since large price changes are typically caused by significant news, which is beyond the scope of our study, we excluded such exceptional days from our data analysis.

	This exceptional handling is anticipated to mitigate selection bias in measuring price impact for large $Q$. For instance, a trader might pause the execution of a buy metaorder if the price goes up unexpectedly fast~\cite{BouchaudText}. Our approach excludes exceptional trading days with sudden price jumps, during which significant implementation bias could exist. 
	
\paragraph*{Appendix D: Stock-level price impact.}
	For a given stock, we defined the daily volatility $\sigma_{\Day}$ as the difference between the daily maximum and minimum prices \cite{PriceRange} (see SM~\cite{SM} for a robustness check using half-day intervals instead of full-day intervals). Also, we defined the daily transaction volume as $V_{\Day}$. Using these characteristic daily parameters, we defined the dimensionless volume and corresponding price impact as 
	\begin{equation}
		\mcQ_k := \frac{Q_k}{V_{\Day}},\>\>\>
		\mcI_k := \frac{I_k}{\sigma_{\Day}}
	\end{equation}
	with the $k$-th metaorder volume $Q_k$ and the corresponding price impact $I_k$. We then studied the average price impact to apply the nonlinear least-square fitting, such that 
	\begin{equation}
		\mcI(Q) := \la \mcI_k \mid \mcQ_k=\mcQ \ra \approx c\mcQ^{\delta}
	\end{equation}
	with the fitting parameters $c$ and $\delta$ at the stock level. Our analysis focuses exclusively on metaorder splitting through market orders. Including metaorder splitting via limit orders would introduce estimation bias in the price impact, as the price impact of limit orders is theoretically negative.
	
	The fitting process is given as follows: First, we extracted the metaorders and the resulting price impacts. As a filter, we focused only on metaorders whose liquidation time horizon---defined as the physical time of metaorder completion---exceeds 60 seconds. This 60-second rule is inspired by the previous research proposing the 120-second rule~\cite{Zarinelli2015}. Second, we introduce logarithmically-evenly spaced bins on the $\mcQ$ axis and take the conditional averages for all bins. Third, we focus only on the bins with sufficient sample sizes (more than 100) and apply the fitting to the conditional averages using the method of relative least squares (for the technical details, see SM~\cite{SM}). We thus obtained $\delta$ and $c$ for all liquid stocks. 
	
	As a robustness check, we also estimated $\delta$ and $c$ using ordinary least squares directly on the scatter plot (without relying on binned averages), and the results remained nearly identical (see SM~\cite{SM}). This confirms that our statistical findings are highly robust, irrespective of the measurement method employed.

\paragraph*{Appendix E: Trader-level price impact.}
	The method for calculating the trader-level price impact is essentially identical to that used for the stock-level price impact. One important remark is that we focused only on active traders. An active trader is defined as one who meets two conditions: First, the total number of their metaorders must be not less than $10^4$. Second, when applying the nonlinear least square fitting to the empirical conditional averages as $\mcI^{(i)}(Q)\approx c^{(i)}\mcQ^{\delta^{(i)}}$, there must be at least 10 bins where the sample size exceeds 100. This filer was introduced to estimate $\delta^{(i)}$ and $c^{(i)}$ with sufficient statistical accuracy. With this definition, the numbers of active traders were 1,293 (415 from Dataset 1 and 878 from Dataset 2). 

\paragraph*{Appendix F: Aggregate scaling plot.}

 The aim of the aggregate scaling plot is to visualise a single average scaling master curve across many stocks. While plotting a few scaling curves, as in Fig.~\ref{fig:DeltaStatistics}(a), is the simplest method to qualitatively check the validity of the square-root scaling, it is unrealistic to plot more than two thousand curves. Therefore, we use the aggregate scaling plot as a practical option to show the qualitative validity of the square-root scaling across all stocks.

	The generating process of the aggregate scaling plot is as follows: First, we obtained the price-impact plots $\mcI(\mcQ)$ for all stocks. We visually checked the validity of the square-root law for all stocks. Second, we applied the fitting $\mcI(\mcQ)\approx c\sqrt{\mcQ}$ with one parameter $c$ for all stocks and obtained the rescaled profiles $\phi(x):=\mcI(\mcQ)/c$ with a variable transformation $x:=\sqrt{Q}$ across all stocks. Third, while we had many $\phi(x):=\mcI(\mcQ)/c$ for all stocks, we calculated their average profile as $\phi_{\rm avg}(x)$. Finally, we plotted $\phi_{\rm avg}(x)$ at the points that are evenly spaced along the logarithmic horizontal axis.

\paragraph*{Appendix G: Numerical simulation.}
	Our nonlinear stochastic model with exact $\delta = 1/2$ is based on our real dataset and random shuffling. (i)~All metaorder execution schedules (in physical time) are identical to those in our real dataset. (ii)~However, the signs of the metaorders are randomly shuffled using our real dataset. (iii)~Furthermore, the price impact of each metaorder is modeled as a Gaussian random variable, following the square-root law on average with the exact half exponent $\delta = 1/2$. The coefficient $c$ and the daily volatility are identical to those in our dataset, but the randomness arises in our model from the shuffling in process~(ii) and the Gaussian random variables in process~(iii). Since our model has an exact half exponent $\delta = 1/2$, it is feasible to evaluate the estimation error of $\delta$ from our real data attributable to the finite sample-size effect, assuming the exact square-root law. 
	

\end{document}


\title{
	{\SM}: \\
		Strict universality of the square-root law in price impact across stocks:\\
		a complete survey of the Tokyo stock exchange
}
\author{Yuki Sato}

\author{Kiyoshi Kanazawa}

\affiliation{Department of Physics, Graduate School of Science, Kyoto University, Kyoto 606-8502, Japan}

\maketitle

\section{Literature review}
	In this section, we briefly review the previous literature to provide enough background knowledge about the square-root price impact law in financial economics and econophysics.

	\subsection{Definition: the square-root price impact law}
		The focus of this paper is the {\it price impact} in financial markets. In financial markets, traders can buy (sell) $Q$ unit of stocks immediately by sending a market order, and the price changes as the market's response. The average price change is called {\it the price impact}, which is defined by
		\begin{equation}
			I(Q) :=  \left\la \eps\left( p(t_{\rm end}) - p(t_{\rm start}) \right) \mid Q\right\ra,
		\end{equation}
		where $\eps$ is the sign of the market order (such as $\eps=+1$ ($\eps=-1$) for the buy (sell) market orders) and $t_{\rm end}$ ($t_{\rm start}$) is the ending (starting) time of the market order (see Chapter 12 of Ref.~\cite{BouchaudText} for the detailed definition of the price impact). Because the buy (sell) market order will increase (decrease) the market price on average, the price impact should be strictly positive $I(Q)>0$. For small perturbation, the linear-response theory is expected to work like physics. Indeed, the price impact is known to be linear for small $Q$, such that $I(Q)\propto \lambda Q$. The coefficient $\lambda$ is called Kyle's lambda after Kyle's pioneering theory~\cite{Kyle}, and many economic models theoretically support the linear law of price impact. 

		\begin{figure*}
			\centering
			\includegraphics[width=180mm]{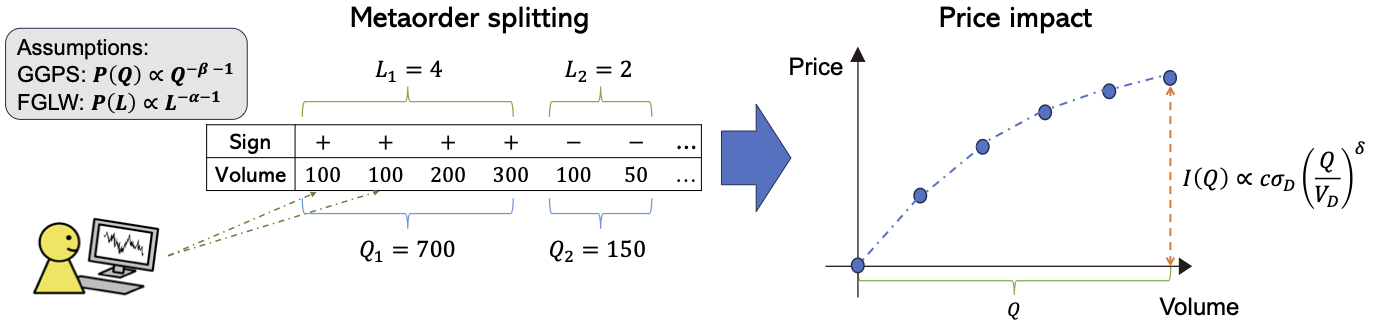}
			\caption{
				Schematic of non-linear price impact. When institutional investors execute large orders (metaorders), they split them into a sequence of child orders. The metaorder splitting is characterised by two quantities: the number of child orders $L$ and the metaorder volume $Q$. These parameters are assumed to obey the power-law distributions, such that $P(L)\propto L^{-\alpha-1}, P(Q)\propto Q^{-\beta-1}$. The resulting average price change is called the {\it price impact}~\cite{BouchaudText} and is defined as the average mid-price change between the beginning and the end of the metaorder. The price impact is a function of the size of the metaorder $Q$ and is denoted as $I(Q)$.
			}
			\label{pic:squarerootlaw}
		\end{figure*}
		For large perturbation, however, such a linear-response theory is experimentally known to be incorrect. Instead, the power-law price impact is observed as the market's nonlinear response, such that
		\begin{align}
			I(Q)\propto c\sigma_{\Day} \left( \frac{Q}{V_{\Day}}\right)^{\delta}\>\>  \mbox{ for large $Q$},
			\label{eq:lreview-SQL-def}
		\end{align}
		where $\delta$ is the characteristic power-law exponent, $V_{\Day}$ is the daily transaction volume, $\sigma_{\Day}$ is the daily volatility, and $c$ is a non-dimensional constant of $O(1)$ (see Fig.~\ref{pic:squarerootlaw} as a schematic). This nonlinear price impact law is called the {\it square-root law} because the typical value of $\delta$ is one half: 
		\begin{equation}
			\delta \approx \frac{1}{2}.
			\label{eq:lreview-delta=1/2}
		\end{equation}
		In this paper, $\delta$ is the central quantity and is presicely measured for solving a long-standing problem in econophysics. 

		We stress that the nonlinear-response relation~\eqref{eq:lreview-SQL-def} is practically important. Indeed, institutional investors (such as hedge funds) often buy (sell) large amounts of stocks when they believe the stock price will go up (down) in future. They typically split the large {\it metaorder} into a sequence of {\it child orders}. This strategic behaviour is called {\it order splitting} and has been a central topic in econophysics~\cite{BouchaudText,LMF_PRE2005,GeneralizedLMF,SatoPRL2023,SatoPRR2023}. Institutional investors practically use the formula~\eqref{eq:lreview-SQL-def} to evaluate the transaction cost during order splitting.

	\subsection{Previous reports on the values of $\delta$}
\begin{table}
  \centering
  \begin{tabular}{|c|c|c|c|c|c|}
			\hline 
				Reference  &  Market  &  Institution & \begin{tabular}{c}Sample size \\ (\# of metaorders)\end{tabular} & Period & $\delta$\\
    \hline  
    Almgren et al.~\cite{Almgren}  
				& US equity market          															 & City group              & 29,509  & 2001-2003  & 0.6 \\
    Moro et al.~\cite{Moro2009}  
				& Spanish Stock Exchange     															& Various traders         & 55,309   & 2001-2004  & 0.7 \\
    Moro et al.~\cite{Moro2009}  
				& London Stock Exchange    															  & Various traders         & 90,393   & 2002-2004  & 0.5 \\
    Zarinelli et al.~\cite{Zarinelli2015}
				& US equity market           															& ANcerno                 & 6,944,883 & 2007-2009  & 0.47 \\
    Bucci et al.,~\cite{Bucci2019a,BucciCoImpact2020,Bucci2020,Bucci2019b}  
				& US equity market           															& ANcerno                 & 8,000,000 & 2007-2009  & 0.5 \\  
    Toth et al.\cite{Toth2011}  
				& Variety of futures contracts 										   & Capital Fund Management & 500,000  & 2007-2010  & 0.5-0.6 \\
				Bacry et al.\cite{Bacry2015}  
				& European equitiy market 															   & European Investment Bank     & 400,000  & 2010  & 0.54 \\
    Bershova et al.\cite{Bershova}  
				& US equity market                          & Alliance Bernstein      & 12,500   & 2009-2011  & 0.5 \\
    Mastromatteo et al.\cite{MastromatteoPRL2014,MastromatteoPRE2014}  
				& Futures contracts market				   & Capital Fund Management & around 1,000,000 & 2008-2012  & 0.4-0.7 \\
    Donier et al.\cite{Donier2015}  
				& Bitcoin market 															& Various traders         & 1,000,000 & 2011-2013  & 0.5 \\
    Brokmann et al.,\cite{Brokmann2015}  
				& \begin{tabular}{c}Equity markets in Europe, \\
        US, Japan, and Australia  
        \end{tabular}  & Capital Fund Management & 1,600,000 & 2011-2013  & 0.6 \\
    Toth et al.,\cite{TothOption}  
				& Option market              															& Capital Fund Management & 450,000  & 2013-2016  & 0.4-0.43 \\
    Said et al.\cite{Said2017}  
				& European equity markets                   & Various traders         & 1,561,505 & 2016-2017  & 0.51 \\
    Said et al.,\cite{Said2021}  
				& Korean options market    & BNP Paribas             & 149,441  & 2016-2018  & 0.53-0.56 \\
    \hline
  \end{tabular}
		\caption{Summary of the previous researches reporting $\delta$. The values typically distribute around $\delta\approx 0.5$, supporting the square-root law. }
  \label{tab:MetaOrder}
	\end{table}
		
		It is not easy to measure $\delta$ because it requires microscopic financial datasets to track traders' behaviour at an individual level. Most of public datasets do not include traders' identifiers; the metaorder size $Q$ and the correspoinding price impact $I(Q)$ cannot be estimated in principle. Due to the scarcity of microscopic datasets, there has been no reliable complete survey of $\delta$ for all the stocks and all the traders. 

		While there has been no complete survey, a small group of researchers measured $\delta$ using their proprietry datasets. Here the proprietry datasets mean that they use trading-log datasets of some specific institutions. For example, the Paris group of Bouchaud is very famous because they manage their own hedge fund (Capital Fund Management) and experimentally measure $\delta$ by their own trading records. Table~\ref{tab:MetaOrder} summarises the previous values of $\delta$ based on such proprietry datasets. The exponent $\delta$ typically distributes around $0.5$, and the nonlinear-response relation~\eqref{eq:lreview-SQL-def} is thus called the square-root law.

	\subsection{Long-standing problem: universal or nonuniversal power-law exponent?}
		We next formulate the research topic in this article. First, we defined the {\it universal (nonuniversal) power-law exponent} as a technical word in traditional statistical physics. We next proceed with the ongoing debate in market microstructure/econophysics regarding the universality argument of the square-root law. We then discuss the reason why this problem has not been solved for a long year from the viewpoint of the difficulty in statistical analysis and data availability. Finally, we clarify the goal of this article together with our solution. 

		\subsubsection{Review: universal power-law exponents in physics}
			In statistical physics, the power-law behaviour is ubiquitously observed, particularly in the context of phase transition and critical phenomena. For example, the ferromagnetic material exhibits the second-order phase transition, and the magnetization $m$ shows the power-law behaviour 
			\begin{equation}
				m \sim 
				\begin{cases}
					|T-T_c|^{\delta} & (T\leq T_c) \\
					0 &  (T>T_c)
				\end{cases}
			\end{equation}
			near the critical temperature $T_c>0$. Here the power-law exponent $\delta$ characterises the phase transition and is known to be determined only by the symmetry and dimension of the target system. 
			
			For example, let us consider the Ising model as the minimal model of the ferromagnetic material. Beyond the upper critical dimension $D\geq 4$, the Landau mean-field theory predicts $\delta = 1/2$, which is exactly correct. For $D=3$, on the other hand, the renormalization-group theory predicts the correct value of $\delta$ that depends on the system's symmetry. 
			
			Such a power-law exponent is called the {\it universal power-law exponent} because it is insensitive to the detail of the microscopic systems; it is not a function of microscopic parameters (such as the lattice constant and coupling constant) but is a function of only global characters of the system (such as the system's symmetry and dimension). The universal power-law exponent is a very good quantity to measure because its value is theoretically very robust; if we want to test some microscopic theory, we should measure this exponent and compare with the theoretical prediction in validating/rejecting the theory.

			On the other hand, if the power-law exponent depends on the details of microscopic parameters of the target systems, the exponent is called the {\it nonuniversal power-law exponent}; it is a function of microscopic parameters $\bm{\theta}:=(\theta_1,\dots, \theta_K)$, such that $\delta = \delta(\bm{\theta})$. The nonuniversal power-law exponent is a fragile quantity to predict because it is sensitive to the details of the microscopic systems. If an exponent $\delta$ is nonuniversal, it means that there exist key microscopic parameters in fixing the value of $\delta$, and precise measurement of such microscopic parameters is inevitable. 

		\subsubsection{Research question of this article}
			Let us go back to our research topic of this paper particularly for market microstructure and econophysics. Our research question is given as follows: 
			\begin{quote}
				Q. Is $\delta$ in the square-root law~\eqref{eq:lreview-delta=1/2} a universal power-law exponent? In other words, is $\delta$ always one-half within statistical errors, independent of the details of financial market microstructure, such as stocks and trading strategies?
			\end{quote}
			To answer this research question, we test two mutually-exclusive hypotheses as follows: 
			\begin{itemize}
				\item $H_0$: $\delta$ is a universal power-law expont and is always one-half within statistical errors. 
				\item $H_1$: $\delta$ is a nonuniversal power-law exponent and is a function of key microscopic parameters $\bm{\theta}$, such that $\delta = \delta(\bm{\theta})$. By measuring $\bm{\theta}$ from microscopic datasets, the correlation should be clearly observed between $\delta$ and $\bm{\theta}$. 
			\end{itemize}

			Finally, we rejected $H_1$ for most of the main scenarios of previous researches, but did not reject $H_0$ in our statistical analyses. We thus present the current best evidence supporting the universality argument of $\delta$. 

		\subsubsection{Review: current theoretical arguments about the square-root law from the universality viewpoint}
			In the literature, there are several microscopic models to explain the origin of the square-root law, and it is a central issue if $\delta$ is a universal exponent or not. For example, the following two theoretical models are famous arguing the nonuniversality of $\delta$: Gabaix-Gopikrishnan-Plerou-Stanley (GGPS) model and Farmer-Gerig-Lillo-Waelbroeck (FGLW) model. 
			
			Both models share the assumptions that (i)~there are the institutional investors splitting large metaorders into sequences of child orders and (ii)~the statistics of the metaorder volume or their run length determines the exponent $\delta$. More precise formulation is given as follows: let us denote the total metaorder volume for one splitting by $Q$ and the total number of child orders (or the run length of one metaorder) by $L$. The distribution of $Q$ and $L$ are assumed to obey the power law: 
			\begin{equation}
				P(Q) \propto Q^{-\beta-1}, \>\>\> P(L) \propto L^{-\alpha-1}
			\end{equation} 
			for large $Q$ and $L$. GGPS and FGLW claim that the key microscopic parameter $\theta$ characterising $\delta(\theta)$ is $\beta$ and $\alpha$, respectively. 

			\begin{itemize}
				\item{\tb{1) Gabaix-Gopikrishnan-Plerou-Stanley (GGPS) model:}} 
					In 2003 and 2006, GGPS claims that the nonlinearity of the price impact originates from the rational inventory-management strategy of market makers in the presence of order splitters~\cite{NatureGabaix, Gabaix}. Because order splitters sometimes buy/sell a huge amount of stocks finally, the inventory risk is high in this setup. They therefore assume that the market maker aims control the inventory risk during their market-making (which is implemented in the utility function as the risk aversion). Finally, they claim that $\delta$ is nonuniversal; $\beta$ is the key microscopic parameter characterising $\delta$ as 
					\begin{equation}
						\delta = \delta(\beta) = \beta - 1, \>\>\> 
					\end{equation}
					where they claim that typical value of $\beta$ is $1.5$, assuming that the fund's asset size distribution follows Zipf's law\footnote{More technically, the exponent $\delta$ is predicted to follow $\delta=\zeta^{-1}\beta-1$ for a general $\zeta$.} $P(F)\propto F^{-\zeta-1}$ with $\zeta=1$. This relationship clearly argues the nonuniversality of $\delta$ because it predicts that $\delta$ can be different from one-half if $\beta$ is not equal to $1.5$.

				\item{\tb{2) Farmer-Gerig-Lillo-Waelbroeck (FGLW) model:}}				
					The FGLW model is a natural extension of the Kyle's model~\cite{FGLW}, assuming the existence of the order-splitters. Market makers attempt to set fair prices inferring the total number of metaorder length from historical order-flow data. As the Nash equilibrium, the price impact is given by a nonlinear profile with the power-law exponet 
					\begin{equation}
						\delta = \delta(\alpha) = \alpha - 1,
					\end{equation}
					where the typical value of $\alpha$ is claimed to be $1.5$. This theory predicts that the key microscopic parameter is $\alpha$, and $\delta$ should deviate from one-half if $\alpha \neq 1.5$.
			\end{itemize}
			The GGPS and FGLW models are the main microscopic models claiminig the nonuniversality of the power-law exponent $\delta$.	On the other hand, there are researchers claiming the universality of the power-law exponent $\delta$. 
			\begin{itemize}			
				\item{\tb{3) Universal square-root-law hypothesis}:}	
					This hypothesis states the power-law exponent $\delta=1/2$ is a universal exponent; in other words, there is no specific microscopic parameter $\bm{\theta}$ controlling $\delta$, and the value of  $\delta$ is robustly $1/2$ for any market and any stock irrelevantly to market microstructure. While there is no established microscopic theory to explain the origin of this universal power-law exponent, recently, the latent-order book model was proposed by Donier et al.~\cite{LatentOrderBook} as a promising microscopic minimal model to understand the universality of $\delta =1/2$.
			\end{itemize}

		\subsubsection{Review: exising data-analytical results and their technical challenges}
			The universal and nonuniversal hypotheses are contradictory to each other, and only one hypothesis should be correct in reality. However, the current situation is controversial from the viewpoint of data analysis. To test these hypotheses, it is necessary to measure $\delta$ for all the stocks and all the traders individually and to draw the scatterplot between $\delta$ and $\alpha$ (or $\beta$). However, there are several technical challenges, which we will explain one by one. 
			
			\begin{itemize}
				\item{\tb{P1) Data availability of microscopic datasets:}} 
					The first challenge is that tracking all the metaorders and measuring their corresponding price impact requires microscopic datasets that track all traders individually. Since such microscopic datasets are rarely available in reality, complete surveys to measure $\delta$ have been lacking so far. 
				
				\item{\tb{P2) Controlling selection biases:}} 
					Controlling selection bias is a very important aspect of statistical analyses but was very difficult in the absence of complete microscopic datasets~\cite{SelectionBias}. The typical design of previous researches has been to use proprietry datasets of some specific companies/institutions (for example, Refs.~\cite{Almgren, Zarinelli2015, Bucci2019a, BucciCoImpact2020, Bucci2020, Toth2011, Bacry2015, Bershova, MastromatteoPRL2014, MastromatteoPRE2014, Brokmann2015, TothOption, Said2021}). However, relying on proprietary datasets means that the statistical results might be biased because it is based neither on a complete survey nor on random sampling. To establish the universality of $\delta$, we should show that the $\delta$ depends neither on the selection of specific traders nor stocks. 
				
				\item{\tb{P3) Large dataset is necessary to control statistical errors within 0.1 for each stock:}} 
					From microscopic datasets, we have to extract $\delta$, $\alpha$, and $\beta$ for each stock for plotting the scatterplots. However, $\delta$, $\alpha$, and $\beta$ are power-law exponents, whose precise measurement requires large sample size for each stock. Also, we have to measure $(\delta, \alpha, \beta)$ for many independent stocks to draw their scatterplot to test the prediction of the GGPS and FGLW models. To provide convincing evidence, we should control the measurement errors of $(\delta, \alpha, \beta)$ within 0.1 for more than 1,000 independent stocks. 

				\item{\tb{P4) Controlling market conditions:}} 
					Many of previous researches mix metaorder datasets for several {\it different stocks} to increase the sample size. They implicitly assume that the heterogeneity in market conditions is completely removed through appropriate non-dimensionalisation. For example, let us express the price impact for stock $s$ as $I_s(Q) = C_sQ^{1/2}$ and assume that the coefficient $C_s$ is always non-dimensionalised by $C_s=c_{s}\sigma_{\Day;s}/V_{\Day;s}^{1/2}$, where $c_s$ is non-dimensional . If this scaling assumption holds completely, we would expect the average {\it across different stocks} for $\mcI(\mcQ):= I_s(Q)/\sigma_{\Day;s}$---the dimensionless price impact conditional on the dimensionless volume $\mcQ:= Q/V_{\Day;s}$---to obey the square-root law. However, this is a strong assumption that requires validation, as mixing different datasets with evident heterogeneity is not desirable in statistics.
			\end{itemize}

			Solving all these problems is challenging, and previous research has not yet conclusively addressed our research question. For example, Bershova and Rakhlin~\cite{Bershova} state that the FGLW prediction is consistent with their empirical observations using a proprietry dataset. On the other hand, Mastromatteo, Toth, and Bouchaud~\cite{MastromatteoPRE2014} shows a scatterplot between $\delta$ and $\gamma$---which is defined as the power-law exponent in the market-order sign autocorrelation function and  a indirect proxy of $\alpha$ because the Lillo-Mike-Farmer model~\cite{LMF_PRE2005,GeneralizedLMF} states the relation $\gamma=\alpha-1$ should hold---using their proprietry dataset. The total number of data points in their scatterplot is ten, which is few, and their estimator is expected to be statistically biased\footnote{
				In our previous publications~\cite{SatoPRL2023,SatoPRR2023}, we found that estimating $\alpha$ is actually feasible from $\gamma$ via the Lillo-Mike-Farmer prediction $\alpha=\gamma+1$, which partially validates the statistical analysis in Ref.~\cite{MastromatteoPRE2014}. However, at the same time, we found that the estimated values of $\alpha$ are shown to be systematically biased along with large statistical errors. Because the standard linear regression formulation assumes that the explanatory variables are measured without statistical errors or biases, statistical interpretation becomes more challenging than usual~\cite{SelectionBias}. Indeed, apparent correlation in scatterplots is known to be generally biased toward zero if its explanatory variable has large measurement error. This phenomenon is called the {\it attenuaion bias} in statistics~\cite{SelectionBias}, which might not be negligible when one uses $\gamma$ as a proxy of $\alpha$.}. 
			However, their scatterplot contradicts the FGLW prediction in the sense that the correlation is not clearly observed between $\delta$ and $\alpha$. In addition, most of all the previous researches are based on propriety datasets and thus selection biases are not well controlled. Thus, convincing and conclusive evidence is necessary to resolve this controversial situation.
			
	\subsection{Goal of this report}
		This paper aims to solve all the problems P1-P4 and to provide convincing and conclusive evidence for the universality of the square-root law. We used a microscopic dataset in the Tokyo Stock Exchange (TSE) market provided by Japan Exchange Group Inc. (JPX) as a complete survey regarding the price-impact law. This dataset covers all the stocks on the TSE at the level of all individual accounts for eight years and thus enables us to track the price impact of each metaorder. The total number of stocks was about one thousand, and the total number of metaorders across all the stocks was approximately $10^8$, which is sufficient as the data size in controlling the measurement errors of $\delta$ within 0.1 for all liquid stocks. Furthermore, we measured the price impact at both market and individual-trader levels as a complete survey, for which we can completely control sampling bias. 
		
		We thus design our statistical analysis as a three steps: 
		\begin{itemize}
			\item Result 1) We measured $\delta$ for {\it all stocks individually} to plot $\delta$'s histogram, supporting the universality hypothesis $\delta=0.5$ on the stock level.
			\item Result 2) We measured $\delta$ for {\it all traders individually} to plot $\delta$'s histogram, supporting the universality hypothesis $\delta=0.5$ on the trader level.
			\item Result 3) We rejected the main predictions of the GGPS and FGLW models, which were previously considered promising in supporting the nonuniversal-exponent hypothesis. 
		\end{itemize}

	\section{Data description and market rule}\label{sec:DataDescription}
		\subsection{Dataset}
			In this research, we use trade and quote data provided by JPX Group Inc., which manages the platform for the TSE. Our dataset contains complete records of order submissions for all the listed stocks on the TSE; i.e., all traders' order submission activities are trackable. In this report, we analysed all liquid stocks, exchange traded funds (ETFs), and real estate investment trusts (REITs) listed on the TSE market from 4 January 2012 to 2 November 2019.
						
			One of the significant advantages of our dataset is that it contains virtual server IDs, which enable us to track metaorder execution. The virtual server identifier (ID) is a unit of account in the TSE. Although it is not completely equivalent to the membership ID, appropriate aggregation of virtual server IDs, {\it the trading desks}~\cite{Goshima2019}, can be an effective proxy of the membership ID. In this report, the trading desk is regarded as the effective membership ID and is called the trader ID for short (for details of the trading desk, see Appendix~\ref{app:sec:tradingdesk} and Refs.~\cite{SatoPRL2023, SatoPRR2023}).
			
		\subsection{Trading rule}
		In this section, we explain the trading rules of the TSE. Understanding these rules is crucial as our data preprocessing is based on them. By briefly describing the trading rules, we aim to provide a clear basis for our data preprocessing approach.

			\subsubsection{Trading system: the {\it arrowhead}}
				Let us describe the trading system at the TSE. The TSE employs an order-matching engine known as {\it arrowhead}, whose performance significantly influences the matching speed. The arrowhead system was updated on 24 September 2015, improving its reaction speed to ten times faster than before. Such a technical update may influence trading behaviour, for example, among high-frequency traders (HFTs) or algorithmic traders. It could also impact our statistical analysis. We have, therefore, divided the dataset into two periods: one from 4 January 2012 to 18 September 2015, and another from 24 September 2015 to 2 November 2019. We refer to the former as Dataset 1 and the latter as Dataset 2.

			\subsubsection{Trading sessions}
\begin{figure*}
					\includegraphics[width=170mm]{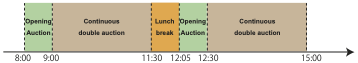}
					\caption{
						The trading sessions at the TSE include three types: (i) the opening auction from 08:00 to 09:00 and from 12:05 to 12:30, (ii) the continuous double auction from 09:00 to 11:30 and from 12:30 to 15:00, and (iii) the closing auction at 11:30 and at 15:00.
					}
					\label{fig:Preprocessing}
				\end{figure*}

				In the TSE, there are three types of trading sessions (see Fig.~\ref{fig:Preprocessing} for a schematic): (i) the opening auction during 08:00-09:00 and 12:05-12:30, (ii) the continuous-double auction during 09:00-11:30 and 12:30-15:00\footnote{
					There was a rule change regarding the continuous-double auction periods from 5 November 2024. The morning session starts from 09:00 to 11:30, and the evening session starts from 12:30 to 15:30.}, 
				and (iii) the closing auction at 11:30 and 15:00. Note that the time zone is set to be Japan; Japan Standard Time (JST) is equivalent to UTC+9. 

				During the opening auction, orders are collected but not immediately executed; they are executed only at the end of the opening auctions, either at 09:00 or 12:30. Conversely, during the continuous double auctions, all orders can be immediately executed through market-order submissions. This paper focuses on studying the order flow during the continuous double auction periods, particularly the price impact of metaorders.

		\subsubsection{Exceptional handling: daily price limit and circuit breaker}
			The TSE has exceptional rules known as the {\it daily price limit} and {\it circuit breaker}, which are aimed at protecting investors from sudden price jumps. In most cases, large price changes are accompanied by significant news. For example, the circuit breaker was triggered at Panasonic Holdings Corporation (ticker: 6752) on 11 January 2012, likely due to an earnings forecast revision. We excluded the days of such exceptional handlings from our data analysis.

		\subsection{Mathematical notations}
			In this subsection, we explain our mathematical notations used to characterise our data. Note that any variable representing sets is written in {\it bold} font, and the others are written in the {\it italic} font. Typically, superscripts with round brackets in any variable represent the trader ID (such as $i$ in $A^{(i)}$), and subscripts represent operating days or stocks (such as $j,s$ in $A_{j,s}$).

			\subsubsection{Extracted variables from the dataset}
			Let us define the variable sets that completely characterise our dataset.
				
\begin{figure*}
				\includegraphics[height=50mm]{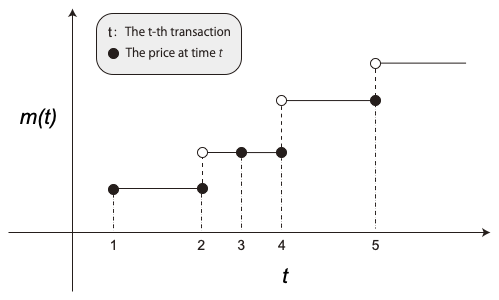}
				\caption{
					The definition of the midprice. In this report, we define the midprice at tick time $t$ as the price just before the $t$-th market-order arrives.
				}
				\label{fig:PriceAtT}
			\end{figure*}

				\begin{itemize}
					\item $\bm{\Omega}_{\Period}$: The set of the operating days. 
				
					\item $\bm{\Omega}_{\Market}$: The set of all stocks. Our dataset is divided into two parts: Dataset 1 and Dataset 2. For our data analysis, each stock in Dataset 1 is considered distinct from any stock in Dataset 2, even if their ticker codes are identical. The total numbers of stocks analysed in each dataset were 942 for Dataset 1 and 1,357 for Dataset 2, respectively.
					
					\item $\bm{\Omega}_{\TR}$: The set includes all trader IDs. Trader IDs are defined for each stock; even if Trader 1 submits orders to two different stocks, such as Stock 1 and Stock 2, we define two separate trader IDs, namely `Trader 1 in Stock 1' and `Trader 1 in Stock 2'. 

					\item $m_{j;s}(t)$: The midprice at tick time $t\in \bm{N}$ on the $j$-th operating day for the $s$-th stock. The tick time $t$ is a positive integer incremented for each transaction. In this report, we define $m_{j;s}(t)$ as the midprice just before the $t$-th market order arrives (see Fig.~\ref{fig:PriceAtT}). The midprice is defined as 
					\begin{align}
						m_{j;s}(t):= \frac{a_{j;s}(t) + b_{j;s}(t)}{2},
					\end{align}
					where $a_{j;s}(t)$ denotes the lowest ask limit price and  $b_{j;s}(t)$ denotes the highest bid limit price. The daily volatility $\sigma_{\Period;j;s}$ is defined as the difference between the highest and lowest daily midprices, and is used as a normalization parameter for the square-root law, according to Parkinson volatility~\cite{PriceRange}:
					\begin{align}
						\sigma_{\Period;j;s} = \max_{t} m_{j;s}(t) - \min_{t} m_{j;s}(t).
					\end{align}

					\item $\mathfrak{t}_{j;s}(t)$: Physical time corresponding to the tick time $t$ on the $j$-th operating day for the $s$-th stock. We measure physical time as the elapsed time [sec] since the continuous double auction starts. We set 09:00:00 (JST) as $0$ [sec] and 15:00:00 as $18,000$ [sec]. 
					
					\item $\eps_{j;s}(t)$: The market-order sign at time $t$ on the $j$-th operating day for the $s$-th stock. The buy (sell) market order is defined as $\eps_{j;s}(t) = +1$ ($\eps_{j;s}(t) =-1$). The number of transactions is defined as $N_{j;s}$ on the $j$-th operating day for the $s$-th stock.
					
					\item $\eps_{j;s}^{(i)}(t)$: The market order sign issued by trader $i \in \bm{\Omega}_{\TR}$ at tick time $t$ on the $j$-th operating day for the $s$-th stock. If trader $i$ did not issue any order at tick time $t$, $\eps_{j;s}^{(i)}(t)$ is set to be zero: $\eps_{j;s}^{(i)}(t)=0$. By definition, an identity holds such that 
					\begin{equation}
						\eps_{j;s}(t) = \sum_{i \in \bm{\Omega}_{\TR}} \eps_{j;s}^{(i)}(t). \label{eq:RelationIndMacro_sign}
					\end{equation}

					\item $v_{j;s}(t)$: The order volume of the $t$-th market order on the $j$-th operating day for the $s$-th stock. We also define normalization parameter of the of $v_{j;s}(t)$ as total transaction volume in the day:
					\begin{align}
						V_{\Period;j;s} = \sum_{t=1}^{N_{j;s}} v_{j;s}(t)
					\end{align}					

					\item $v_{j;s}^{(i)}(t)$: the volume of the $t$-th market order issued by trader $i$ on the $j$-th operating day for the $s$-th stock. If trader $i$ did not issue any order at tick time $t$, $v_{j;s}^{(i)}(t)$ is set to be zero: $v_{j;s}^{(i)}(t)=0$. By definition, an identity holds such that 
					\begin{equation}
						v_{j;s}(t) = \sum_{i \in \bm{\Omega}_{\TR}} v_{j;s}^{(i)}(t). \label{eq:RelationIndMacro_volume}
					\end{equation}
			\end{itemize}

			For our statistical analysis, one stock can be completely characterised by the set of these quantities: 
			\begin{align*}
				\bmGamma_{s}:=\{
					\eps_{j;s}(t),
					m_{j;s}(t),
					\eps^{(i)}_{j;s}(t),
					v_{j;s}(t),
					v_{j;s}^{(i)}(t)
					\}_{i\in\bm{\Omega}_{\TR}, j\in\bm{\Omega}_{\Period}}
			\end{align*}
			This characteristics $\bmGamma_{s}$ is defined for each stock $s\in\bm{\Omega}_{\Market}$.

			\subsubsection{Other important quantities}
				\begin{figure*}
					\centering
					\includegraphics[width=180mm]{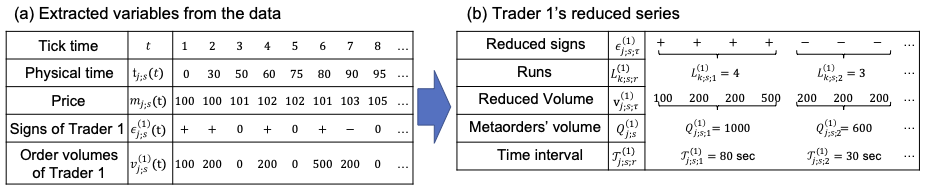}
					\caption{
						Schematic example of the definitions of metaorders and their related quantities.  
					}
					\label{fig:MathNotation}
				\end{figure*}
				We next define several important quantities related to $\bmGamma_{s}$ (see Fig.~\ref{fig:MathNotation}):
				\begin{itemize}
					\item $\{\eps^{(i)}_{j;s;\tau}\}_{\tau}$: Reduced order-sign sequences on the $j$-th operating day for the $s$-th stock, by removing zeros from the original order-sign sequences $\{\eps_{j;s}^{(i)}(t)\}_{t\in \bm{N}}$. The total number of the market orders submitted by trader $i$ is denoted by $N^{(i)}_{j;s;\rm MO}:=|\{\eps^{(i)}_{j;s;\tau}\}_{\tau}|$. Note that $|\{A_i\}_{i}|$ implies the number of the total elements of the set $\{A_i\}_i$. For example, $|\{A_i\}_{i=1,\dots,K}|=K$ with a positive integer $K$.

					\item $\{L^{(i)}_{j;s;r}\}_{r}$: Runs sequence based on the reduced-sign sequences $\{\eps^{(i)}_{j;s;\tau}\}_{\tau}$ for trader $i \in \bm{\Omega}_{\TR}$. For a given reduced-sign sequence, the {\it runs} are defined as as the numbers of adjacent equal elements in the sequence $\{\eps^{(i)}_{j;s;\tau}\}_{\tau}$; e.g., $L_1^{(1)}=4$ and $L_2^{(1)}=3$ for $\{\eps^{(i)}_{j;s;\tau}\}_{\tau}=\{++++---+...\}$. In this paper, a metaorder from trader $i$ is defined based on a run; market orders are considered child orders if they successively share the same sign and are placed by the same trader. The set of these child orders is regarded as a metaorder. Additionally, metaorders are defined within a single day; when the market closes at 15:00:00 (JST), we consider the metaorder to be completed. The total number of metaorders submitted for stock $s$ is $M_{s}:=\sum_{i\in\Omega_{\TR}, j\in\Omega_{\Period}} |\{L^{(i)}_{j;s;r}\}_r|$. Also, the total number of metaorders submitted by trader $i$ for stock $s$  is $M_{s}^{(i)}:=\sum_{j\in \bm{\Omega}_{\Day}}|\{L^{(i)}_{j;s;r}\}_r|$.

					\item $\{v^{(i)}_{j;s;\tau}\}_{\tau}$: Reduced order-volume sequences of trader $i$ on the $j$-th operating day for the $s$-th stock market, by removing zeros from the original order-sign sequences $\{v_{j;s}^{(i)}(t)\}_{t\in \bm{N}}$.

					\item $\{Q^{(i)}_{j;s;r}\}_{r}$: Total volume of the $r$-th metaorder by trader $i$ on the $j$-th operating day for the $s$-th stock. 

					\item $\{\mathcal{T}^{(i)}_{j;s;r}\}_{r}$: Physical time interval during which the $r$-th metaorder by trader $i$ was completed on the $j$-th operating day for the $s$-th stock. The unit is seconds. In this paper, $\mathcal{T}^{(i)}_{j;s;r}$ is called the {\it liquidation time horizon}. 
 				\end{itemize}

			\subsubsection{Notation for probability theory and indicator function}
				The probability density function (PDF) or the normalised histogram characterise the probability that a stochastic variable $X$ satisfies $X \in [x,x+\dd x)$ as $P(x)\dd x$. For a given series $\{x_k\}_k$, the empirical PDF or the normalised histogram can be defined by 
				\begin{equation}
					P(x) := \frac{1}{\left|\{x_k\}_k\right|}\sum_{k} \delta(x-x_k),
				\end{equation}
				which satisfies the normalisation condition $\int P(x)\dd x=1$. Here, $\delta(x)$ is Dirac's delta function satisfying the following conditions: $\delta(x)=0$ for $x\neq 0$, $\delta(0)=\infty$, and $\int_{-\infty}^\infty f(x)\delta(x)\dd x=f(0)$ for any smooth function $f(x)$. In this article, the normalised histogram is referred to as the histogram for short.

				We also define the indicator functions $\mathbbm{1}(A=a)$ and $\mathbbm{1}(A \in [b,c))$, such that 
				\begin{equation}
					\mathbbm{1}(A=a) = \begin{cases}
						1 & \mbox{if $A=a$} \\
						0 & \mbox{if $A\neq a$} 
					\end{cases}, \>\>\>
					\mathbbm{1}(A \in [b,c)) = 
					\begin{cases}
						1\>\>\>\mbox{if } A  \in [b,c) \\
						0\>\>\>\mbox{otherwise}
					\end{cases}
				\end{equation}
				with stochastic variable $A$ and real numbers $a$, $b$, and $c$. 

				In this work, we define three types of averages, particularly to distinguish the average for the TSE dataset from those for our statistical model. To avoid possible confusion, we summarize them as follows: 
				\begin{itemize}
					\item For the TSE dataset, the empirical average of $A$ across all stocks is written as $\la A\ra$.
					\item For our numerical statistical model, the Monte Carlo average of $A$ for a specific stock $s$ is written as $\dla A_s\dra:=(1/N_{\rm MC})\sum_{l=1}^{N_{\rm MC}} A_{s;l}$, where $N_{\rm MC}$ is the total number of Monte Carlo simulations and $A_{s;l}$ is the value of $A$ for stock $s$ on the $l$-th Monte Carlo iteration.  
					\item For our numerical statistical model and for one specific Monte Carlo iteration, the average of $A$ across all stocks is written as $\overline{A_l}:=(1/|\bm{\Omega}_{S}|)\sum_{s\in\bm{\Omega}_{S}} A_{s;l}$, where $A_{s;l}$ is the value of $A$ for the specific stock $s$ on the $l$-th iteration.
				\end{itemize}

			\subsubsection{Data extraction rule}\label{sec:DataPreprocessing}
				We describe the our data extraction rule and exceptional handling regarding the selection of stocks and periods. For our statistical analysis, we extracted only an appropriate part of our data in control various statistical/structural noise and biases. 
				\begin{itemize}
					\item{\tb{Sample-size control:}}
						We focused only on sufficiently liquid stocks where the total number of observations is large regarding metaorders. If the total number of metaorders is insufficient, such that $M_{s}\leq 10^5$ for the stock index $s$, the stock $s$ was excluded from our statistical analysis. This rule was introduced to control measurement errors and bias regarding the sample size. 

					\item{\tb{Exceptional rule regarding the price limit or circuit breaker:}}
						Trading days were excluded if the exceptional rule of the price limit or the circuit breaker was satisfied. This exceptional handling was introduced to remove the effect of significant news that are out of scope of our research. 

					\item{\tb{Exceptional rule regarding the liquidity crisis:}}
						Trading days were excluded if there were extraordinarily large market orders that sweeped all the limit orders. We regarded such trading days as outliers that are out of scope of our research. 
				\end{itemize}

	\section{Empirical price impact on the stock level}\label{sec:PriceImpact-Market}
		In this section, we describe the statistical properties of the stock-level price impact. 

		\subsection{Definition of the price impact}
			\begin{figure*}
				\includegraphics[width=140mm]{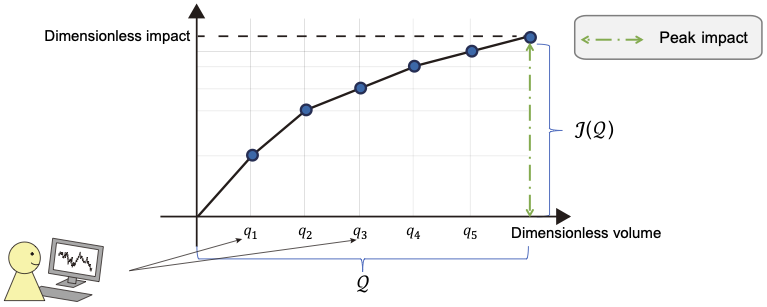}
					\caption{
						Schematic of price impact conditional on the size of the metaorder $Q$. The price impact refers to the average difference between the initial and final prices resulting from the metaorder $Q$ (illustrated by the green line), where $q_k$ represents the cumulative volumes up to the $k$-th child order (i.e., $q_1<q_2<\dots<q_6=Q$). Note that this measurement method of the price impact is technically called {\it peak impact}. 
					}
					\label{fig:TotalImpact-PathImpact-Schematic}
			\end{figure*}
			In this article, we quantify the price impact based on the {\it peak impact} (see Fig.~\ref{fig:TotalImpact-PathImpact-Schematic} and Ref.~\cite{BouchaudText}): The peak impact $\mcI_{s}(\mcQ)$ is defined for the stock $s$ as the average price change between the starting and ending times of a metaorder with total size $\mcQ$:
			\begin{align}
				\mcI_{s}(\mcQ):= \left\la \mcI_{j;s;r}^{(i)} \mid \mcQ_{j;s;r}^{(i)} = \mcQ \right\ra 
			\end{align}
			with dimensionless impact and dimensionless volume defined by 
			\begin{equation}						
				\mcI_{j;s;r}^{(i)}:= \frac{1}{\sigma_{\Day;j;s}}
				\left\{m_{j;s}(t_{\End;j;s;r}^{(i)}+1)-m_{j;s}(t_{\Start;j;s;r}^{(i)})\right\}\eps(t_{\Start;j;s;r}^{(i)}), \>\>\> 
				\mcQ_{j;s;r}^{(i)} := \frac{Q}{V_{\Day;j;s}},
			\end{equation}
			where $t_{\Start;j;s;r}^{(i)}$ and $t_{\End;j;s;r}^{(i)}$ represent the tick times corresponding to the beginning and end of the $r$-th metaorder by the $i$-th trader on the $j$-th operating day for the $s$-th stock. 
			
			Note that the conditional average $\la A_{j;s;r}^{(i)} \mid \mcQ_{j;s;r}^{(i)}=\mcQ \ra$ refers to the conditional empirical average across all the traders, all the metaorders, and all the trading days, such that 
			\begin{equation}
				\left< A_{j;s;r}^{(i)} \mid \mcQ_{j;s;r}^{(i)}=\mcQ \right> := \frac{\sum_{j,r,i}A_{j;s;r}^{(i)} \mathbbm{1}(\mcQ_{j;s;r}^{(i)}=\mcQ)}{\sum_{j,r,i} \mathbbm{1}(\mcQ_{j;s;r}^{(i)}=\mcQ)}, 
			\end{equation}
			for any quantities $A_{j;s;r}^{(i)}$. When we add some filters (such as the 60-second rule) in the following subsections, the indicator function should be replaced appropriately. In this article, the dimensionless peak impact is simply referred to as the price impact when its meaning is clear from the context.

		\subsection{Statistical results for the stock-level price impact}~\label{sec:TotalImpact}
			In this subsection, we describe the statistical properties of the price impact for each stock. We measure the price impact for the $s$-th stock based on the following set:
			\begin{align}
				\mclM_{s}:= 
					\left\{
						(
							\mcI_{j;s;r}^{(i)},  
							\mcQ_{j;s;r}^{(i)}
						)
						  \mid
					\mathcal{T}^{(i)}_{j;s;r}\geq 60 \mbox{ [sec]}
				\right\}_{
					j\in\bm{\Omega}_{\Period},
					i\in\bm{\Omega}_{\TR},
					r\in[1,N^{(i)}_{j;s;\rm MO}]
				},
			\end{align}
			where the 60-second rule was introduced based on the 120-second filter used in previous research~\cite{Zarinelli2015}, because the square-root law is expected to be observed with sufficiently slow metaorder execution. This filter is applied for the measurement of the conditional average $\la A_{j;s;r}^{(i)} \mid \mcQ_{j;s;r}^{(i)}=\mcQ \ra$.  

			 \subsubsection{Empirical price-impact profile $\mcI_{s}(\mcQ)$}
				\begin{figure*}
					\includegraphics[width=180mm]{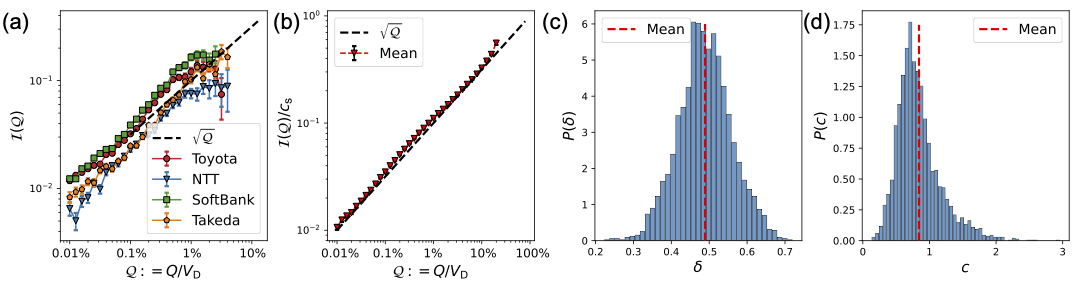}
					\caption{
						Summary statistics for the stock-level price impact. 
						(a)~Average price impact profile for the four specific stocks from Dataset 2: The red line represents Toyota Motor Corp. (ticker 7203), the blue line represents Nippon Telegraph and Telephone (NTT) Corp. (ticker 9432), the green line represents SoftBank Group Corp. (ticker 9984), and the orange line represents Takeda Pharmaceutical Company Limited (ticker 4502). Error bars represent the standard errors of the means (SEMs).
						(b)~aggregate scaling plot, a technical method to visualize the average scaling priofile across all stocks (see Appendix~\ref{app:sec:aggscaling}), reveals that the price impact conforms to the square-root law $\mcI_{\Total}(\mcQ)\propto c \sqrt{\mcQ}$ for all stocks.
						(c)~Empirical histogram of the power-law exponent $\delta_s$ across all stocks $s \in \bm{\Omega}_{\Market}$. The exponent $\delta_s$ distributes sharply around $0.5$ with $\la\delta\ra= 0.489$ with its SEM of 0.0015 and $\sigma_\delta:= \sqrt{\la (\delta- \la\delta\ra)^2\ra}= 0.071$. The size of the standard deviation $\sigma_\delta$ is attributable to the finite sample-size effect, according to our simple price-series model (see Appendix~\ref{app:sec:simulation}).
						(d)~Empirical histogram of the prefactor $c_{s}$ across all stocks $s \in \bm{\Omega}_{\Market}$. The typical value of the prefactor $c_{s}$ is of $O(1)$, with $\la c_{s}\ra = 0.842$. 
					}
					\label{fig:MarketPriceImpact}
				\end{figure*}
				To develop our intuition, we present Fig.~\ref{fig:MarketPriceImpact}(a), which plots the price impact as a function of the metaorder volume $\mcQ$ for four specific stocks: Toyota Motor Corp. (with ticker 7203; represented by the red line), Nippon Telegraph and Telephone (NTT) Corp. (with ticker 9432; blue line), SoftBank Group Corp. (with ticker 9984; green line), and Takeda Pharmaceutical Company Limited (with ticker 4502; orange line) from Dataset 2. Figure~\ref{fig:MarketPriceImpact}(a) shows that the square-root law was actually observed for these four stocks: 			
				\begin{align}
					\mcI_{s}(\mcQ) = c_{s}\mcQ^{\delta_s},\>\>\>\delta_s \approx 0.5. \label{eq:peakimpact}
				\end{align}
				where all of $\mcI_{s}(\mcQ)$, $c_{s}$ and $\delta_s$ are measured for each stock $s \in \{7203, 9432, 9984, 4502\}$ from Dataset 2. 

				We next present the aggregate scaling plot (see Appendix~\ref{app:sec:aggscaling} for detailed implementation) as Fig.~\ref{fig:MarketPriceImpact}(b), which displays the average price-impact profile across all stocks. The purpose of the aggregate scaling plot is to illustrate the price-impact profile for all stocks collectively, since it is impractical to depict individual price-impact curves for each stock as demonstraded in Fig.~\ref{fig:MarketPriceImpact}(a). The aggregate scaling plot in Fig.~\ref{fig:MarketPriceImpact}(b) demonstrates that the square-root law~\eqref{eq:peakimpact} is applicable on average across all stocks\footnote{Note that we visually inspected the individual price-impact curves for each stock to confirm the applicability of the square-root law~\eqref{eq:peakimpact} across all stocks.}. We thus conclude that it is reasonable to assume that the price impact can be approximated by a power-law type nonlinear function.

			\subsubsection{Empirical histogram of $\delta_s$ and $c_s$}				
				Furthermore, we plot the empirical distributions of $\delta_s$ and $c_s$ across all the stocks as Fig.~\ref{fig:MarketPriceImpact}(c) and (d), respectively: 
				\begin{equation}
					P(\delta):= \frac{1}{|\bm{\Omega}_{\Market}|}\sum_{s \in \bm{\Omega}_{\Market}}\delta(\delta - \delta_s),\>\>\> 
					P(c):= \frac{1}{|\bm{\Omega}_{\Market}|}\sum_{s \in \bm{\Omega}_{\Market}}\delta(c - c_{s}).
				\end{equation}
				The parameters $\delta_s$ and $c_{s}$ are estimated by the method of relative least squares (RLS) for binned averages, assuming the power-law fitting function $I_{s}(\mcQ)=c_{s} \mcQ^{\delta_s}$ for each stock $s$ (see Appendix~\ref{app:sec:measurement} for technical details). 

				Let us discuss the implications from Fig.~\ref{fig:MarketPriceImpact}(c) and (d). Firstly, Figure~\ref{fig:MarketPriceImpact}(c) shows that $\delta$ distributes around $\delta=0.5$ with a sharp peak. The average and standard deviation of $\delta_s$ are given by 
				\begin{equation}
					\la \delta_s \ra = 0.489\pm 0.0015, \>\>\> \sigma_\delta:=\sqrt{\la (\delta_s - \la\delta_s\ra)^2\ra}= 0.071,
				\end{equation}
				where the standard error of the mean (SEM) was estimated to be $0.0015$. 
				The estimated errorbar is given by $\dla \overline{\sigma_{\delta}}\dra\approx 0.063$, as will be discussed soon.
				This result supports the universality hypothesis of the square-root law $\delta=0.5$ within this statistical errors. Secondly, Fig.~\ref{fig:MarketPriceImpact}(d) shows that the dimensionless coefficient $c_{s}$ is typically of $O(1)$ for all stocks. Indeed, we have $\la c \ra \approx 0.842$. This result is consistent with the previous reports~\cite{Donier2015,TothOption,Said2017,Said2021,BouchaudText}. 
				
				Note that our statistical results were robust to variations in the fitting method, liquidation time horizon, and trading period dependencies (see Appendix~\ref{app:sec:robcheck}). For instance, the histograms of $\delta$ and $c$ obtained using ordinary least squares (OLS) directly on the scatterplot (without relying on binned averages) were nearly identical to those obtained through the RLS estimation.

			\subsubsection{Numerical evaluation of the finite sample-size effect}
\begin{figure}
					\includegraphics[width=90mm]{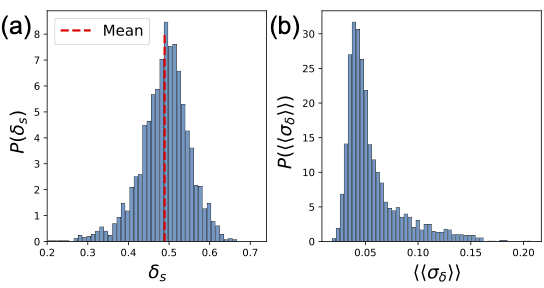}
					\caption{
						Errorbar estimation based on our numerical statistical model. We estimate $\delta_{s;l}$ for all stocks $s\in \bm{\Omega}_{S}$ on the $l$-th Monte Carlo iteration, and repeat this process for $N_{\rm MC}=100$ times. (a)~Histogram of $\delta_{s}$ across all stocks for one specific Monte Carlo iteration. The histogram has a clear single peak around the mean value $\overline{\delta_{s,l}}\approx 0.5$. (b)~Histogram of errorbars $\dla \sigma_{\delta}\dra_{s}$ for all individual stocks $s\in \bm{\Omega}_S$. The errorbar $\dla \sigma_{\delta}\dra_{s}$ for stock $s$ is measured through the Monte Carlo simulation, in the frequentist sense.  
					}
					\label{fig:SM:Hist-Errobars-simulation}
				\end{figure}

				We next estimate the errorbar of $\delta$ attributable to the finite sample-size effect. For this purpose, 
				we considered a simple price-series model with exact square-root impacts by randomly shuffling the order signs (i.e., buys or sells) of metaorders for each stock, while keeping the metaorder execution schedule unchanged from our dataset. See Appendix~\ref{app:sec:simulation} for the motivation to construct a plausible statistical model (Appendix~\ref{app:sec:simulation-motivation}), the detailed implementation (Appendix~\ref{app:sec:simulation-model}), and the final results (Appendix~\ref{app:sec:simulation-results}). While $\delta_s=1/2$ is exact for this numerical model by construction, we repeatedly generated price series and measured $\delta_s$ under the same condition as our dataset. 
				See Fig.~\ref{fig:SM:Hist-Errobars-simulation}(a), where the $\delta$'s histogram  across all stocks for one specific Monte Carlo iteration is illustrated with clear peak around $\overline{\delta_{s;l}}\approx 0.5$ consistently with the histogram~\ref{fig:MarketPriceImpact}(c) based on the TSE dataset. 
				
				Finally, we numerically obtain 
				\begin{equation}
					\dla\overline{\delta_s}\dra=0.489\pm 0.0013, \>\>\> \dla \overline{\sigma_\delta} \dra:=\sqrt{\dla \overline{(\delta_s - 1/2)^2}\dra}= 0.063.
				\end{equation}
				Here, let us write the value of $A$ as $A_{s;l}$ for the stock $s$ and the $l$-th Monte Carlo iteration based on our numerical statistical model (instead of the TSE dataset). The average $\dla A_s \dra$ represents the Monte Carlo average of $A_{s;l}$ across all $l$, assuming the same sample size as in our dataset. In addition, $\overline{A_s}$ represents the cross-sectional average of $A_s$ across all stocks for one specific Monte Carlo iteration (see Appendix~\ref{app:sec:simulation} for the details). More technically, we define them by 
				\begin{equation}
					\dla A_s \dra:=\frac{1}{N_{\rm MC}}\sum_{l=1}^{N_{\rm MC}}A_{s;l}, \>\>\> 
					\overline{A_{s;l}}:=\frac{1}{|\bm{\Omega}_{\rm S}|}\sum_{s\in \bm{\Omega}_{\rm S}}A_{s;l}, \>\>\> N_{\rm MC}=100.
				\end{equation}				
				Because the following relations hold
				\begin{equation}
					|\la\delta_s\ra-\dla\overline{\delta_s}\dra|\ll \dla \overline{\sigma_\delta} \dra, \>\>\> \dla \overline{\sigma_\delta} \dra \approx \sigma_\delta,
				\end{equation}
				the universal power-law hypothesis $\delta=0.5$ is maintained within statistical errors on the basis of our numerical null model\footnote{
					Technically, there are two candidates of errorbars: $\dla \sigma_{\delta} \dra$ is one candidate adopted in our main text. Another is the simple mean of the $\delta's$ SEM across individual stocks (see Fig.~\ref{fig:SM:Hist-Errobars-simulation}): 
					\begin{equation}
						\overline{\dla \sigma_{\delta} \dra} := \sum_{s\in \bm{\Omega}_{\rm S}}\frac{\dla \sigma_{\delta} \dra_s}{|\bm{\Omega}_{\rm S}|} \approx 0.057, 
					\end{equation} 
					where $\bm{\Omega}_{\rm S}$ is the set of all stocks and $\dla \sigma_{\delta} \dra_s$ is the $\delta$'s SEM for the specific stock $s \in \bm{\Omega}_{\rm S}$. $\dla \overline{\sigma_{\delta}} \dra$ is slightly different from $\overline{\dla \sigma_{\delta} \dra}$ because these two averages do not commute. However, they are  connected by 
					\begin{equation}
						\dla \overline{\sigma_{\delta}} \dra^2
						=\frac{1}{N_{\rm MC}}\sum_{l=1}^{N_{\rm MC}}\frac{1}{|\bm{\Omega}_{\rm S}|}\sum_{s\in \bm{\Omega}_{\rm S}} \left(\delta_{s;l}-\frac{1}{2}\right)^2
						=\frac{1}{|\bm{\Omega}_{\rm S}|}\sum_{s\in \bm{\Omega}_{\rm S}} \dla \sigma_{\delta} \dra^2_s = \overline{\dla \sigma_{\delta} \dra^2_s}.
					\end{equation}
					In the main text, we employed $\dla \overline{\sigma_{\delta}} \dra$ as our errorbar instead of $\overline{\dla \sigma_{\delta} \dra_s}$, because $\dla \sigma_{\delta} \dra$ is directly related to the cross-sectional dispersion $\sigma_{\delta}$ on the TSE dataset. Due to the Cauchy-Schwarz inequality, we can prove $\dla \overline{\sigma_{\delta}} \dra \geq \overline{\dla \sigma_{\delta} \dra}$, consistent with our results. 
				}. See also Fig.~\ref{fig:SM:Hist-Errobars-simulation}(b) for the histogram of errobars for individual stocks, where the errobar $\dla \sigma_\delta\dra_s$ for the specific stock $s\in \bm{\Omega}_S$ and its histogram $P(\dla \sigma_\delta\dra)$ are defined by 
				\begin{equation}
						\dla \sigma_{\delta} \dra_s :=  \sqrt{\frac{1}{N_{\rm MC}}\sum_{l=1}^{N_{\rm MC}} \left(\delta_{s;l} - \frac{1}{2} \right)^2}, \>\>\> 
						P\left(\dla \sigma_{\delta} \dra\right) := \frac{1}{|\bm{\Omega}_{\rm S}|}\sum_{s\in \bm{\Omega}_{\rm S}}\delta\left(\dla \sigma_{\delta} \dra-\dla \sigma_{\delta} \dra_s\right).
				\end{equation}

			\subsection{Main message of Sec.~\ref{sec:PriceImpact-Market}}
				In Sec.~\ref{sec:PriceImpact-Market}, we found that the square-root law with half exponent $\delta\approx 1/2$ for all stocks. The deviation from one-half, $|\delta - 1/2|$, can be attributed to the finite sample-size effect in measuring $\delta_s$. We thus concluded that the universal-exponent hypothesis was supported within statistical errors (at least, not rejected on the TSE) in our stock-level analysis.

	\section{Empirical price impact on the individual trader level}~\label{sec:Trader}
		\begin{figure*}
			\includegraphics[width=150mm]{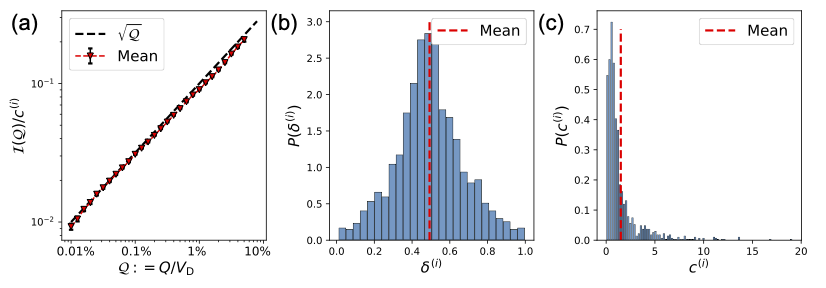}
			\caption{
					Summary statistics for the trader-level price impact. 
					(a)~aggregate scaling plot of the trader-level price impact. The rescaled price impact confirms to the square root law $\mcI_{\Total}(\mcQ)\propto c_{s}^{i}\sqrt{\mcQ}$.
					(b)~Empirical distribution of the power-law exponent $\delta_s^{(i)}$ for the range $\delta_s^{(i)} \in[0,1]$. Note that we observe a few outliers within $\delta \not \in[0,1]$, which seemingly originate from fitting errors. The average value of $\delta_{s}^{(i)}$ was given by $\la\delta_{s}^{(i)}\ra=0.493$ with its SEM of 0.0050, which was very near to $0.5$. 
					(c)~Empirical histogram of the prefactor $c_{s}^{(i)}$. Typically, the prefactor $c_{s}^{(i)}$ is of $O(1)$ with the average $\la c_{s}^{(i)}\ra=1.501$.
			}
			\label{fig:Trader}
		\end{figure*}
		Since the universal-exponent hypothesis was supported on the stock level, let us proceed with the verification/rejection of the universal-exponent hypothesis on the individual trader level in this section. 

		\subsection{Definition of the price impact for active traders}
			In this section, we focus on active traders (metaorder splitters) which are defined by  
			\begin{equation}
				\bm{\Omega}_{\AT;s} := \{i \mid i \in \bm{\Omega}_{\TR}, M_{s}^{(i)} \geq 10^4, N_{{\rm bin};s}^{(i)}>10\},
			\end{equation}
			where $N_{{\rm bin};s}^{(i)}$ is the total number of bins that were valid for parameter estimation\footnote{We applied a filter to control the estimation errors to define valid bins; see Appendix~\ref{app:sec:measurement} for the details.}. The total number of the active traders was 1,293 (415 from Dataset 1 and 878 from Dataset 2). For an active trader $i \in \bm{\Omega}_{\AT;s}$, we define the dimensionless price impact as follows: The price impact $\mcI_{s}^{(i)}(\mcQ)$ is defined for an active trader $i\in \bm{\Omega}_{\AT;s}$ and for the stock $s$ as the average price change between the starting and ending times of a metaorder with total size $\mcQ$:
			\begin{align}
				\mcI_{s}^{(i)}(\mcQ):= 
				\left\la 
					\mcI_{j;s;r}^{(i')} 
					\mid \mcQ_{j;s;r}^{(i')} = \mcQ, \>\>\> i'=i 
				\right\ra.
			\end{align}
			Here, the conditional average $\la A_{j;s;r}^{(i')'} \mid \mcQ_{j;s;r}^{(i')}=\mcQ ,\>\>\> i'=i \ra$ refers to the conditional empirical average for an active trader $i \in \bm{\Omega}_{\AT}$ across all the metaorders, and all the trading days, such that 
			\begin{align}
				\left< A_{j;s;r}^{(i')} \mid \mcQ_{j;s;r}^{(i')}=\mcQ, \>\>\> i'=i \right> &:= \frac{\sum_{j,r}A_{j;s;r}^{(i)} \mathbbm{1}(\mcQ_{j;s;r}^{(i)}=\mcQ)}{\sum_{j,r} \mathbbm{1}(\mcQ_{j;s;r}^{(i)}=\mcQ)}
			\end{align}
			for any quantities $A_{j;s;r}^{(i)}$. When we add some filters (such as the 60-second rule) in the following subsections, the indicator function should be replaced appropriately.

		\subsection{Statistical results for the trader-level price impact}	
			To measure the price impact at the level of individual traders, we extracted the following information from our dataset for an active trader $i \in \bm{\Omega}_{\AT}$:
			\begin{equation}
				\mclM^{(i)}_{s}:= 
				\left\{
					(
						\mcI_{j;s;r}^{(i)}, 
						\mcQ_{j;s;r}^{(i)}
					)
					\mid 
					\mathcal{T}^{(i)}_{j;s;r}\geq 60 \mbox{ [sec]}
					\right\}_{
						j\in\bm{\Omega}_{\Period},
						r\in[1,N^{(i)}_{j;s;\rm MO}]
					}.
			\end{equation}
			We focused on only metaorders whose liquidation time horizon is not less than 60 seconds. Based on the information $\mclM^{(i)}_{s}$, we studied the price impact $\mcI_{s}^{(i)}(\mcQ)$ for each active trader with the scaling assumption: 
			\begin{equation}
				\mcI_{s}^{(i)}(\mcQ) = c_{s}^{(i)}\mcQ^{\delta_{s}^{(i)}}, \>\>\> 
				i \in \bm{\Omega}_{\AT}.
			\end{equation}

		\subsubsection{Empirical price-impact profile $\mcI_{s}^{(i)}(\mcQ)$ and histograms of $\delta^{(i)}_s$ and $c^{(i)}_s$}
			Figure~\ref{fig:Trader}(a) presents the aggregate scaling plot of the trader's price impact, demonstrating that the square-root law holds at the trader level, despite the heterogeneity of trading strategies. We visually inspected the impact profiles of all individual traders and confirmed that the square-root law holds for each of them before generating the aggregate scaling plot.

			Figure~\ref{fig:Trader}(b) shows the empirical distribution of $\delta_{s}^{(i)}$, such that $P(\delta) \propto \sum_{i,s} \delta(\delta-\delta_{s}^{(i)})$. This figure illustrates that the exponent $\delta$ sharply distributed around $\delta=0.5$ even at the individual-trader level. Indeed, we obtained the average and the standard deviation of $\delta_s^{(i)}$:
			\begin{equation}
				\la \delta_{s}^{(i)}\ra = 0.493\pm 0.0050, \>\>\> \sigma_{\delta^{(i)}_s}=0.177.
			\end{equation}
			As a reference, we present Fig.~\ref{fig:Trader}(c) as the histogram of the coefficient $c_{s}^{(i)}$, such that $P(c) \propto \sum_{i,s} \delta(c-c_{s}^{(i)})$. A sharp peak was found around $c_{s}^{(i)} \sim 1$, with an average of $\la c_{s}^{(i)}\ra=1.501$. 

		\subsubsection{Numerical evaluation of the finite sample-size effect}
			To investigate the origin of the observed standard deviation $\sigma_{\delta^{(i)}_s}$, we performed numerical simulation of a simple null model following the exact square-root law (see Appendix~\ref{app:sec:simulation}). Based on the method of relative least squares, we numerically obtain  
			\begin{equation}
				\dla\delta^{(i)}_s \dra = 0.521 \pm 0.0048, \>\>\> 
				\dla \sigma_{\delta^{(i)}_s}\dra= 0.169.
			\end{equation}
			Since $\sigma_{\delta^{(i)}_s}\approx \dla \sigma_{\delta^{(i)}_s}\dra$, the level of the standard deviation in our real dataset is  at the same level as the finite sample-size effect in our numerical null model. Therefore, we conclude that the square-root law holds at the trader level within statistical errors.
								
		\subsection{Main message of Sec.~\ref{sec:Trader}}
			In Sec.~\ref{sec:Trader}, we studied the price impact at the level of individual traders and measured $\delta_{s}^{(i)}$. We confirmed that the empirical histogram for $\delta_{s}^{(i)}$ exhibits sharp peak around $\delta_{s}^{(i)}=0.5$, with means $\la\delta_{s}^{(i)}\ra=0.493$. The size of the standard deviation $\sigma_{\delta^{(i)}_s}$ can be attributed to the finite sample-size effect, according to our numerical null model. This result supports the robustness of the square-root law, suggesting the universality of $\delta$ regardless of trading strategies, even at the level of individual traders.

	\section{Empirical test of theoretical predictions of the nonuniversal hypotheses}~\label{sec:FGLW-GGPS}
		\begin{figure*}
			\includegraphics[width=180mm]{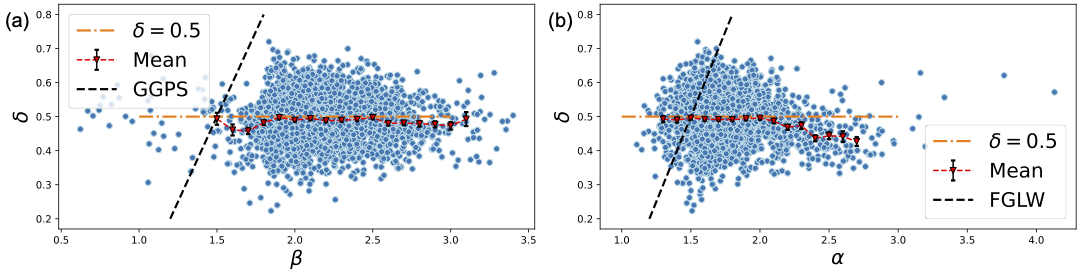}
			\caption{
				Scatterplots examining the GGPS and FGLW predictions, (a)~$\delta=\beta-1$ and (b)~$\delta=\alpha-1$. No correlation was found, and the GGPS and FGLW predictions were rejected.
			}
			\label{fig:FGLW-GGPS-Delta}
		\end{figure*}
		In this section, we test the theoretical predictions of the FGLW and GGPS models, two main models supporting the nonuniversal hypothesis. These two models predict that the power-law exponent of the price impact $\delta$ should depend on the metaorder distributions, such that 
		\begin{align}
			\mbox{GGPS model prediction: }\>\>\> \delta&=\beta-1,\label{eq:GGPS}\\
			\mbox{FGLW model prediction: }\>\>\> \delta&=\alpha-1.\label{eq:FGLW}
		\end{align}
		The exponents $\beta$ and $\alpha$ represent the power-law exponents for the metaorder volume distribution ($P(Q)\propto Q^{-\beta-1}$) and the metaorder length distribution ($P(L)\propto L^{-\alpha-1}$), respectively. Using our dataset, we measured $\beta$ and $\alpha$ through Clauset's algorithm~\cite{Clauset}. 	
	
		To test the predictions~\eqref{eq:GGPS} and \eqref{eq:FGLW}, it is sufficient to create two scatterplots: one between $\beta$ and $\delta$, and the other between $\alpha$ and $\delta$. Figure~\ref{fig:FGLW-GGPS-Delta}~(a) and \ref{fig:FGLW-GGPS-Delta}~(b) are the scatterplots between $\beta$ and $\delta$, and between $\alpha$ and $\delta$, respectively. Both figure shows the absence of correlation, rejecting the theoretical predictions in~\eqref{eq:FGLW} and \eqref{eq:GGPS}. Interested readers should see Appendix~\ref{app:sec:robcheck}, where the robustness check is performed for the scatterplots. 

\appendix
	
	\section{Trading desk constitution}\label{app:sec:tradingdesk}
		\begin{table}
			\begin{tabular}{c|c|c|c}
					\hline 
						Virtual server ID  &  Order ID  &  Order type & Trading desk \\
				\hline \hline
						V1  &  O1   &  Limit order  & T1 \\
						V2  &  O1   &  Cancel order & T1 \\
						V3  &  O2   &  Limit order  & T2 \\
						V4  &  O4   &  Limit order  & T3 \\
						\vdots  &  \vdots   &  \vdots  & \vdots \\
			\end{tabular}
			\caption{
				How to define the trading desks. While the lifecycle of a single order ID is typically managed by a single virtual server ID, there are exceptional cases where multiple virtual server IDs are involved in the lifecycle of one order. For example, consider the case where virtual server ``V1" submits a limit order, and then virtual server ``V2" cancels the same limit order under the same order ID ``O1." In this instance, we infer that the same trader operates both virtual servers, V1 and V2, and we assign a unified identifier, ``T1," as the effective trader ID. This trader ID is referred to as the trading desk.
			}
		\label{tab:trading-desk}
		\end{table}
		The {\it trading desk} is an effective trader ID that is defined when we find several virtual servers are possessed by the same membership. Let us describe the technical process to define the trading desks. 
		
		Virtual server IDs are not technically equivalent to membership IDs, which are defined at the level of corporate accounts. Each virtual server ID has a limit on the number of submissions within a fixed time interval. Some corporations possess multiple virtual server IDs to circumvent this submission-speed limit. Indeed, we sometimes observe exceptional cases where multiple virtual server IDs are involved in the lifecycle of a single order ID. For example, a limit order with the order ID ``O1" might be submitted from virtual server ID ``V1" and later canceled from virtual server ID ``V2." In such cases, it is evident that both virtual server IDs ``V1" and ``V2" are shared by the same membership (see Table~\ref{tab:trading-desk}).

		The trading desk concept was introduced to address this issue. We assign an effective and unified trader ID so that virtual server IDs ``V1" and ``V2" are associated with the same effective trader ID, ``T1." This effective trader ID, ``T1," is referred to as a {\it trading desk}, as was proposed in Ref. \cite{Goshima2019}. We defined trading desks by checking all the order lifecycle across all the stocks for each dataset. 

	\section{Measurement of the price impact}\label{app:sec:measurement}
		\begin{figure}
			\centering
			\includegraphics[width=180mm]{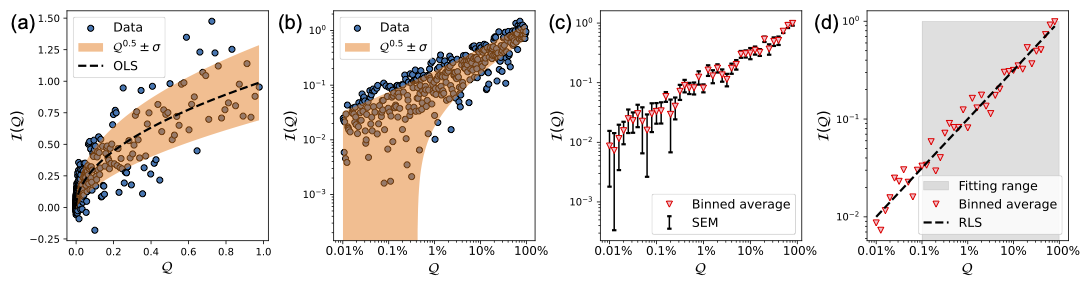}
			\caption{
			Schematic of the RLS fitting based on numerical simulations. 
			(a)~Schematic scatterplot on normal scales illustrating the non-uniform distribution of datapoints and the presence of heteroscedasticity, which directly impacts OLS fitting. 
			(b)~Scatterplot on a log-log scale. 
			(c)~Binned averages displayed on the log-log scale, where the averages are uniformly distributed along the logarithmic $\mcQ$ scale, and the heteroscedasticity is reduced through the averaging process.
			(d)~The parameters $\delta$ and $c$ are estimated via RLS fitting on the binned averages. The contributions of all data points are uniformly weighted on the logarithmic $\mcQ$ scale, enhancing the robustness of the RLS fitting.
			}\label{fig:regression-schematic}
		\end{figure}

		In this Appendix, we describe our statistical methods to estimate the power-law exponent $\delta$ in the price impact based on RLS.

		\subsection{Algorithm for the price impact measurement}

		The details of the statistical method used to measure $(\delta,c)$ in the price impact at the stock level are described below. Note that the same process can be straightforwardly applied to the analysis of price impact at the individual-trader level.
			
			Assuming the necessary and sufficient information $\mclM_{s}:= \left\{(\mcI_{j;s;r}^{(i)}, \mcQ_{j;s;r}^{(i)}) \right\}_{j, i, r}$, we apply the following steps (see Fig.~\ref{fig:regression-schematic} as a schematic):
			\begin{itemize}
				\item{\tb{Step 1. Binning:}} Let us define specific points $\{\bar{\mcQ}_{k}\}_{k=0,1,\dots,60}$ that are evenly distributed on a logarithmic scale:
				\begin{equation}
					\mcQ = \bar{\mcQ}_{k} := 10^{-3+k\Delta} \mbox{ for $k=0, 1,\dots 60$}. \>\>\> \Delta=0.05. 
				\end{equation}
				We calculate the average price impact $\mcI_{s}(\mcQ):=\left<\mcI_{j;s;r}^{(i)} | \mcQ_{j;s;r}^{(i)}=\mcQ\right>$ at the points $\{\bar{\mcQ}_{k}\}_{k}$, such that 
				\begin{align}
					\mcI_{s}(\bar{\mcQ}_{k}) = \frac{1}{N_k}\sum_{j,i,r}\mcI_{j;s;r}^{(i)}\mathbbm{1}(\mcQ_{j;s;r}^{(i)} \in [10^{-\Delta/2}\bar{\mcQ}_{k},10^{\Delta/2}\bar{\mcQ}_{k})),\>\>\>
					N_k=\sum_{j,i,r} \mathbbm{1}(\mcQ_{j;s;r}^{(i)} \in [10^{-\Delta/2}\bar{\mcQ}_{k},10^{\Delta/2}\bar{\mcQ}_{k})).
				\end{align}

				\item{\tb{Step 2. Fitting:}} Based on the empirical conditional average $\mcI_{s}(\bar{\mcQ}_{k})$ at $\{\bar{\mcQ}_{k}\}_{k}$, we estimate the power-law exponent $\delta_s$ and the prefactor $c_{s}$ using the method of RLS: 
				\begin{align}
					\mcI_{\rm model;s}(\mcQ) := c_{s} \mcQ^{\delta_s}, \>\>\> 
					\min_{c_{s},\delta_{s}} \sum_{k=0}^{60}\mathbbm{1}(N_k > 10^2)\left|\frac{\mcI_{s}(\bar{\mcQ}_{k})-I_{\rm model}(\bar{\mcQ}_{k})}{I_{\rm model}(\bar{\mcQ}_{k})}\right|^2,
				\end{align}
				where we applied a filter, focusing on bins with more than 100 data points. We define the number of such valid bins used for the parameter estimation as $N_{{\rm bin};s}:=\sum_{k=0}^{60}\mathbbm{1}(N_k > 10^2)$, and was used to define active traders to control the statistical errors.
			\end{itemize}

			\subsection{Theoretical comparison}
			While our statistical results were almost identical between the RLS and OLS estimations (see Appendix~\ref{app:sec:robcheck}), let us theoretically compare these estimations:  
			\begin{itemize}
				\item{\tb{Ordinary least squares for the scatterplots}:} 
					OLS is one of the most standard statistical methods for estimating model parameters. Notably, when the model assumptions hold, OLS estimators typically have the consistency for an infinite sample size. Given the large sample size of our dataset, OLS was expected to perform reasonably well. 					
					
					However, we noticed three delicate issues when applying the OLS estimation for our dataset: (i)~The datapoints are not uniformly distributed along the logarithmic $\mcQ$ scale. In fact, many datapoints are concentrated at smaller $\mcQ$, resulting in a non-uniform contribution to the loss function along the logarithmic $\mcQ$ axis. (ii)~Heteroscedasticity is clearly present\footnote{
						Naively, the variance is expected to increase as $\sqrt{\mcQ}$ by considering price diffusion, but the data showed that the variance increased with $\mcQ^{a}$ with $a\simeq 0.3$.},
					with the variance increasing as $\mcQ$ grows. This implies that each datapoint contributes to the loss function non-uniformly along the logarithmic $\mcQ$ axis. For an illustration of issues (i) and (ii), see Fig.~\ref{fig:regression-schematic}(a). Additionally, (iii)~while the power-law exponent $\delta$ represents the linear slope of the average price impact on the log-log scale, the OLS estimation does not directly correspond to such an intuitive interpretation. Even in the absence of issues (i) and (ii), the datapoints still contribute to the loss function non-uniformly along the $\mcQ$ axis. 
					
					To illustrate the issue~(iii), let us rewrite the OLS loss function based on the relative errors: 
					\begin{align}
						\mathcal{L}(x,\mcI(\mcQ)) 
						= \sum^{n}_{k=1} \left( x - \mcI(\mcQ) \right)^2
						= \sum^{n}_{k=1} \underbrace{\mcI(\mcQ)}_{\rm weight}\underbrace{\left( \frac{x - \mcI(\mcQ)}{\mcI(\mcQ)} \right)^2}_{\rm relative\> error},
					\end{align}
					where $\mcI(\mcQ)=c_0 \mcQ^{\delta}$. Notably, the relative error is more important in log-log scale plots than the absolute errors. In the OLS fitting over the range $Q\in[10^{-4},10^{-2}]$, for example, the datapoints with $\mcQ\sim O(10^{-2})$ are weighted ten times more heavily than those in $\mcQ\sim O(10^{-4})$, regarding relative errors. 
					
					These three issues highlight that, while OLS theoretically performs well with an infinite sample size, we were unsure if it works well for finite sample sizes due to its inability to uniformly control the contributions of data points along the logarithmic $\mcQ$ scale. 			
				
				\item{\tb{Relative least square fitting to binned averages}:} 
					In contrast, the RLS fitting applied to binned averages is theoretically robust for finite sample sizes from this perspective. Issues (i) and (ii) are mitigated by using binned averages, assuming the bins are uniformly distributed on the logarithmic $\mcQ$ scale (see Fig.~\ref{fig:regression-schematic}(c) for a schematic illustration). Furthermore, the RLS estimation effectively assigns a uniform weight to the relative errors across all value of $\mcQ$:
					\begin{align}
						\mathcal{L}(x,f(\mcQ)) 
						&= \sum^{n}_{k=1} \left( \frac{x - f(\mcQ)}{f(\mcQ)} \right)^2.
					\end{align}
					Thus, the RLS estimation is expected to mitigate the issue (iii). Note that the consistency of this estimation method is reasonably expected, as the binned averages should converge to the true values with an infinite sample size. Indeed, we numerically confirmed that the estimated values were consistent for our theoretical null model (see Appendix~\ref{app:sec:simulation}).
			\end{itemize}

			By considering these theoretical aspects, we studied the RLS estimation to the binned averages, in addition to the conventional OLS estimation. Since both results were almost identical, we trust the statistical robustness of our main results. 
			
	\section{aggregate scaling plot}\label{app:sec:aggscaling}
		\begin{figure}
			\centering
			\includegraphics[width=180mm]{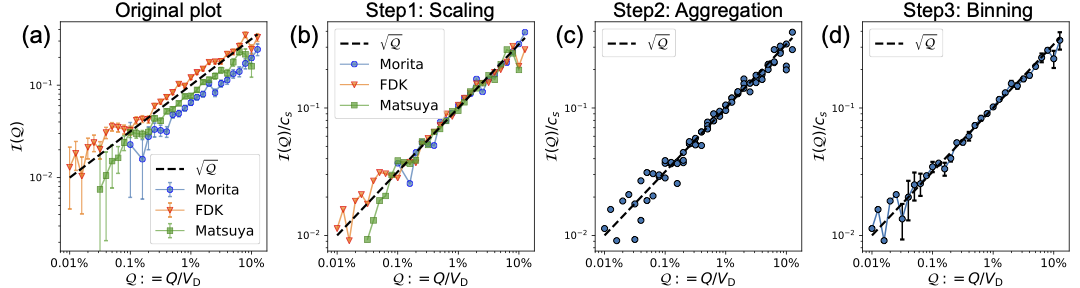}
			\caption{
				Schematic example of the aggregate scaling plots: (a)~Average price impact profile for the three specific stocks from Dataset 2: Morita Holdings Corp. (blue, ticker 6455, $s=1$), FDK Corp. (orange, ticker 6955, $s=2$), and Matsuya Co. LTD. (green, ticker 8237, $s=3$), where $s$ represents the stock index. While this data agrees with the square-root law $\mcI_s(\mcQ)=c_s\sqrt{\mcQ}$ on average, the three coefficient $\{c_s\}_{s=1,2,3}$ are not identical. (b)~This heterogeneity is removed by 
				applying the rescaling $\phi(x)=\mcI_s(\mcQ(x))/c_{s}$ with $x:=\sqrt{\mcQ}$. (c)~Next, the average $\phi_{\rm avg}(x)$ is obtained across all the stocks. (d)~Finally, the averaged profile $\phi_{\rm avg}(x)$ is plotted with bins evenly spaced on a logarithmic scale.
			}
			\label{fig:agg_scaling_proc}
		\end{figure}	
		This Appendix describes the {\it aggregate scaling plots}, which visually display a single master curve after rescaling across all the stocks (or all the traders). The idea is straightforward: (i) Assuming the validity of a scaling function, we rescale both the horizontal and vertical axes to confirm that the data aligns with a single master curve. (ii) This process is repeated for many stocks (or traders), and we take the average across all stocks (or traders) to summarise all the scaling plots into one figure. We use aggregate scaling plots because it is impractical to display an excessive number of individual scaling plots, though we have confirmed the validity of the scaling functions for all stocks (or individual traders).
	
		The detailed process for the stock-level analysis is described below (the same process can be straightforwardly applied to individual-trader-level analysis): As a working hypothesis, we assume that the price impact function in the $s$-th stock follows a power-law form:  
		\begin{equation}
			\mcI_{s}(\mcQ) = c_{s} \mcQ^{1/2}
		\end{equation}
		If this working hypothesis is correct, the following single master curve should apply to all stocks:
		\begin{equation}
			\phi(x) = \frac{\mcI_{s}(\mcQ)}{c_{s}} = x, \>\>\> x:= \mcQ^{1/2},	\label{eq:scaling}
		\end{equation}
		where $c_{s}$ is measured by the method of relative least squares. If this single scaling applies to all the stocks, we take the average for each bins across all the stocks. See Fig. 1 for a schematic of this statistical processing. We plot $\phi_{\rm avg}(x)$ as the aggregate scaling plot.

	\section{Robustness check}\label{app:sec:robcheck}
		The robustness of our findings are discussed in this Appendix. 

		\subsection{Robustness check with half-day (session-based) aggregation}

			\begin{figure*}
				\includegraphics[width=110mm]{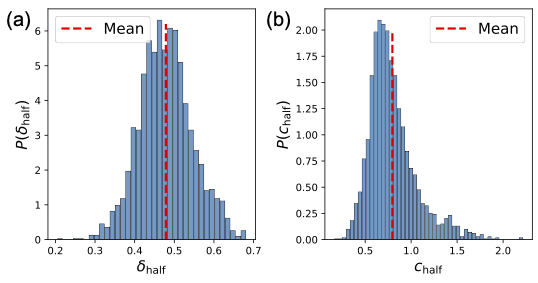}
				\caption{
					Robustness check of the estimated $\delta_{{\rm half}}$ and $c_{{\rm half}}$ based on the half-day (session-based) aggregation.
					(a)~Histogram of $\delta_{{\rm half}}$ with a mean $\la\delta_{{\rm half}}\ra=0.479$.
					(b)~Histogram of the prefactor $c_{{\rm half}}$ with a mean $\la c_{{\rm half}}\ra =0.796$. 
				}
				\label{fig:SessionBySession}
			\end{figure*}
			To ensure the reliability of our findings, we conducted a robustness check on our results with respect to the aggregation timescale. In the main text, we define statistical quantities on a daily basis, such as the metaorder volume and length distributions, the daily volatility, and the daily total volume. Instead of such daily analyses, we also re-examined the results on a half-day basis. More specifically, given that the TSE has two sessions (morning and evening) each day, we used a single session as the unit of aggregation. Figure~\ref{fig:SessionBySession} presents the results of this half-day analysis, demonstrating the robustness of our findings. 	

	\subsection{Robustness check regarding the non-dimensionalization of $\mcQ$ and $\mcI$}
		In the main text, we define the dimensionless variables $\mcQ:=Q/V_{\Day}$ and $\mcI:=I/\sigma_{\Day}$ using $V_{\Day}$ and $\sigma_{\Day}$, the total transaction volume and volatility on the same day, respectively. In this subsection, we explore the results obtained when using $V_{\Day}$ and $\sigma_{\Day}$ from the previous day.
		\begin{figure*}
			\includegraphics[width=110mm]{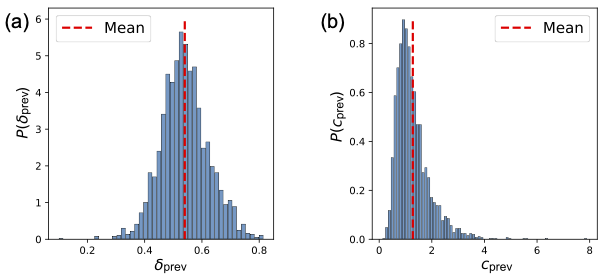}
			\caption{
					Histograms of (a)~$\delta_{{\rm prev}}$ and (b)~$c_{{\rm prev}}$ using  the previous day's total transaction volume $V_{\Day}$ and volatility $\sigma_{\Day}$ for non-dimensionalization: $\mcQ:=Q/V_{\Day}$ and $\mcI:=I/\sigma_{\Day}$. The means are $\la\delta_{{\rm prev}}\ra=0.540$ and $\la c_{{\rm prev}}\ra=0.918$. 
			}
			\label{fig:PrevDayInfo}
		\end{figure*}
		Figure~\ref{fig:PrevDayInfo} shows histograms of $\delta_{{\rm prev}}$ and $c_{{\rm prev}}$, with means $\la\delta_{{\rm prev}}\ra=0.540$ and $\la c_{{\rm prev}}\ra= 0.918$, respectively. These results support the SRL and demonstrate robustness to the choice of dimensionless variables.

		\subsection{Robustness check based on the OLS estimation for $\delta_s$}	
			\begin{figure*}
				\includegraphics[width=180mm]{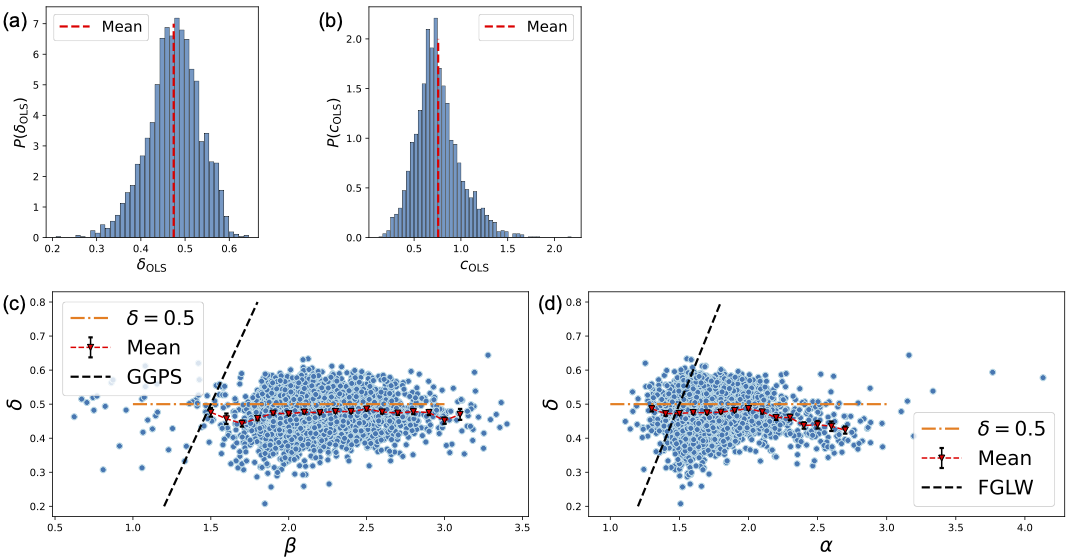}
				\caption{
						Robustness check by estimating $\delta_{\rm OLS}$ and $c_{\rm OLS}$ by the nonlinear OLS. 
						(a)~Histogram of $\delta_{\rm OLS}$ with $\la\delta_{\rm OLS}\ra=0.474$. 
						(b)~Histogram of $c_{\rm OLS}$ with $\la c_{\rm OLS}\ra= 0.757$. 
						(c, d) Scatterplots testing the GGPS and FGLW prediction using $\delta_{{\rm OLS}}$, demonstrating no correlation. All the results are almost identical with those based on the RLS fitting to the binned averages, showing the robustness of our analysis.
				}
				\label{fig:robustnesscheck-OLS}
			\end{figure*}			
			This subsection provides a robustness check by estimating $\delta_{\rm OLS}$ and $c_{\rm OLS}$ using OLS,complementing the main analysis which uses RLS on binned averages. Figures~\ref{fig:robustnesscheck-OLS}~(a) and (b) display histograms of the OLS estimates, $\delta_{\rm OLS}$ and $c_{\rm OLS}$, with means $\la\delta_{\rm OLS}\ra=0.474$ and $\la c_{\rm OLS}\ra= 0.757$. Figure~\ref{fig:robustnesscheck-OLS}(c) shows the scatterplot of $\delta_{{\rm OLS}}$ against $\beta$, while Fig.~\ref{fig:robustnesscheck-OLS}(d) shows the scatterplot of $\delta_{{\rm OLS}}$ against $\alpha$, showing no correlation. These results demonstrate the robustness of our main results to the choice of estimator

		\subsection{Two-fold cross validation regarding the price impact}		
			\begin{figure*}
				\includegraphics[width=180mm]{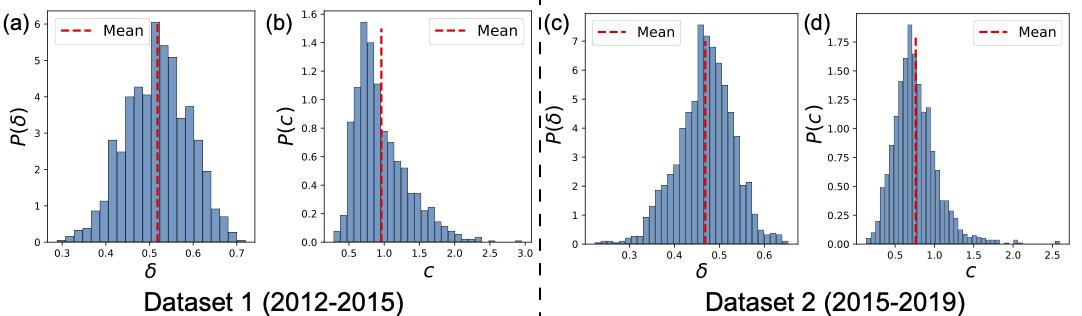}
				\caption{
					Two-fold cross validation regarding the exponent $\delta_s$ and coefficient $c_s$ using Dataset 1 and Dataset 2. 
					(a, b)~Empirical histogram for Dataset 1 (from 2012 to 2015): $\delta_s$ with $\la\delta_s\ra=0.518$ in Fig.~(a) and $c_{s}$ with $\la c_{s}\ra=0.960$ in Fig.~(b). 
					(c, d)~Empirical histogram for Dataset 2 (from 2015 to 2019): $\delta_s$ with $\la\delta_s\ra=0.469$ in Fig.~(c) and $c_{s}$ with $\la c_{s}\ra=0.761$ in Fig.~(d).
				}
				\label{fig:PriceImpact-CV}
			\end{figure*}
			We plot Fig.~\ref{fig:PriceImpact-CV} as the histograms of the measured parameters for $\{(\delta_s,c_{s})\}_{s\in \bm{\Omega}_{\Market}}$, for Dataset 1 (from 2012 to 2015, Fig.~\ref{fig:PriceImpact-CV}(a, b)) and Dataset 2 (from 2015 to 2019, Fig.~\ref{fig:PriceImpact-CV}(c, d)). These figures serve as a two-fold cross-validation and are consistent with Fig.~\ref{fig:MarketPriceImpact}(c,d), supporting the robustness of our results.
	
		\subsection{Robustness check of the exponent $\delta_s$ with respect to the liquidation time horizon}
			\begin{figure*}
				\includegraphics[width=180mm]{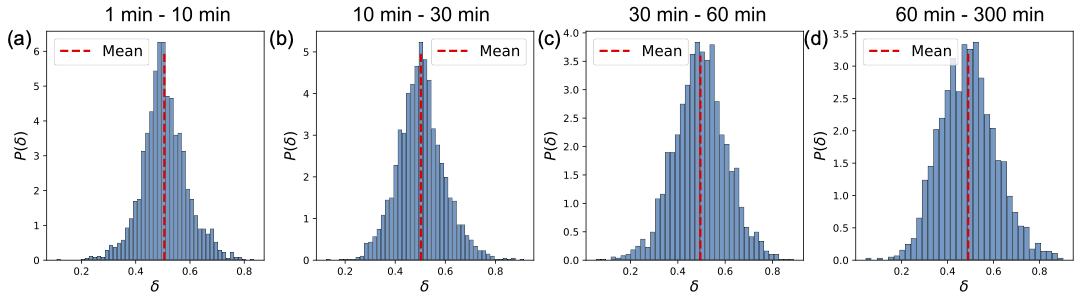}
				\caption{
					Robustness check of the exponent $\delta_s$ by changing the liquidation time horizon $\mathcal{T}^{(i)}_{j;s;r}$. We plotted the empirical histograms of  $\delta_s$ for various time horizons for $\delta\in[0,1]$: (a)~$\mathcal{T}^{(i)}_{j;s;r} \in [ 1 \mbox{ min}, 10 \mbox{ min})$ with $\la \delta_s\ra =0.506$, 
					(b)~$\mathcal{T}^{(i)}_{j;s;r} \in [10 \mbox{ min}, 30 \mbox{ min})$ with $\la \delta_s\ra =0.502$,
					(c)~$\mathcal{T}^{(i)}_{j;s;r} \in [30 \mbox{ min}, 60 \mbox{ min})$ with $\la \delta_s\ra =0.495$, and 
					(d)~$\mathcal{T}^{(i)}_{j;s;r} \in [60 \mbox{ min}, 300 \mbox{ min})$ with $\la \delta_s\ra =0.490$. All these figures have a peak around $\delta_s\approx 0.5$, suggesting the robustness of our results.
				}
				\label{fig:TimeScaleDependencies}
			\end{figure*}
			In this subsection, we assess the robustness of the price-impact analyses in relation to the liquidation time horizon (i.e., the total time required to complete a metaorder), as previous research has suggested the possibility that the price-impact profile might depend on the liquidation time horizon~\cite{RamaCont}. 
			
			Let us define four sets $\{\mclM_{s;k}\}_{k=1,2,3,4}$ as subsets of $\mclM_{s}$, categorized by the liquidation time horizon:
			\begin{subequations}\label{eq:subsets-TimeDependencies}
				\begin{equation}
					\mclM_{s;k}:= 
						\left\{
							\left(
								I_{j;s;r}^{(i)}, 
								Q_{j;s;r}^{(i)} 
							\right)
						\mid 
						\mathcal{T}^{(i)}_{j;s;r} \in \bm{\Omega}_k
					\right\}_{
						j\in\bm{\Omega}_{\Period},
						i\in\bm{\Omega}_{\TR},
						r\in[1,N^{(i)}_{j;s;\rm MO}]
					}
				\end{equation}
				with the four filters
				\begin{equation}
					\bm{\Omega}_1 := [1 \mbox{ min}, 10 \mbox{ min}), \>\>\>
					\bm{\Omega}_2 := [10 \mbox{ min}, 30 \mbox{ min}), \>\>\>
					\bm{\Omega}_3 := [30 \mbox{ min}, 60 \mbox{ min}), \>\>\>
					\bm{\Omega}_4 := [60 \mbox{ min}, 300 \mbox{ min}).
				\end{equation}
			\end{subequations}
			Using these four sets $\{\mclM_{s;k}\}_{k=1,2,3,4}$, we measured four types of $\delta_s$ to created four corresponding histograms, as shown in Fig.~\ref{fig:TimeScaleDependencies}. As an additional filter, we excluded stocks with a total number of metaorders not greater than $10^4$, such that $|\mclM_{s;k}|\leq 10^4$. These figures suggest that the empirical distributions $\delta$ are insensitive to the liquidation time horizon, supporting the robustness of our main result.

		\subsection{Consistency with previous studies regarding the crossover from the linear to the square-root laws}
			\begin{figure*}
				\includegraphics[width=110mm]{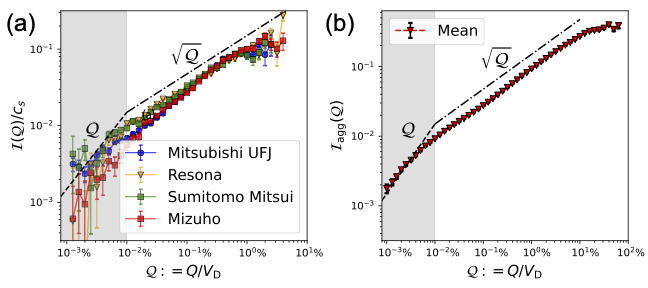}
				\caption{
					Crossover from the linear to the square-root laws. 
					(a)~Average price impact $\mcI(\mcQ)$ as a function of metaorder volumes $\mcQ$ for four bank-related stocks: Mitsubishi UFJ Financial Group, Inc. (blue, ticker 8306), Resona Holdings, Inc. (orange, ticker 8308), Sumitomo Mitsui Financial Group, Inc. (green, ticker 8316), and Mizuho Financial Group, Inc. (red, ticker 8411). For small $\mcQ$, we observe the linear price-impact regime. 
					(b)~Average price impact $\mcI(\mcQ)$ as a function of metaorder volumes $\mcQ$ simply aggregated across all the stocks. 
				}
				\label{fig:PriceImpact-SmallMetaorder}
			\end{figure*}
			Let us check the consistency of our work with previous studies~\cite{Zarinelli2015,Bucci2019a}, where the crossover from the linear law (for small $\mcQ \ll \mcQ^*$) to the square-root law (for large $\mcQ\gg \mcQ^*$) is observed with the characteristic threshold $\mcQ^*$. While we focus only on the nonlinear price-impact regime in the main text, we illustrate the average price impact $\mcI(\mcQ)$ even for the small volume regime $\mcQ \ll \mcQ^*$ in Figure~\ref{fig:PriceImpact-SmallMetaorder}. Specifically, Fig.~\ref{fig:PriceImpact-SmallMetaorder}(a) shows the price impact for four bank stocks. Figure~\ref{fig:PriceImpact-SmallMetaorder}(b) shows the price impact aggregated across all stocks, based on 
			\begin{align}
				\mclM_{\rm agg}:= 
					\left\{
						(
							\mcI_{j;s;r}^{(i)},  
							\mcQ_{j;s;r}^{(i)}
						)
						  \mid
					\mathcal{T}^{(i)}_{j;s;r}\geq 60 \mbox{ [sec]}
				\right\}_{
					j\in\bm{\Omega}_{\Period},
					s\in\bm{\Omega}_{\Market},
					i\in\bm{\Omega}_{\TR},
					r\in[1,N^{(i)}_{j;s;\rm MO}]
				}.
			\end{align}
			Note that Fig.~\ref{fig:PriceImpact-SmallMetaorder}(b) is not an aggregated scaling plot, defined in Appendix~\ref{app:sec:aggscaling}; it is based on the simple binned average, aggregated across all stocks, by following the previous studies~\cite{Zarinelli2015,Bucci2019a}. Finally, we observe the crossover from the linear law for small $\mcQ\ll \mcQ^{*}$ to the square-root law for large $\mcQ\gg \mcQ^{*}$ with the threshold $\mcQ^{*}\approx 10^{-4}$, consistently with Refs.~\cite{Zarinelli2015,Bucci2019a}.
		
		\subsection{Two-fold cross validation regarding the GGPS and the FGLW predictions}
		\begin{figure*}
			\includegraphics[width=180mm]{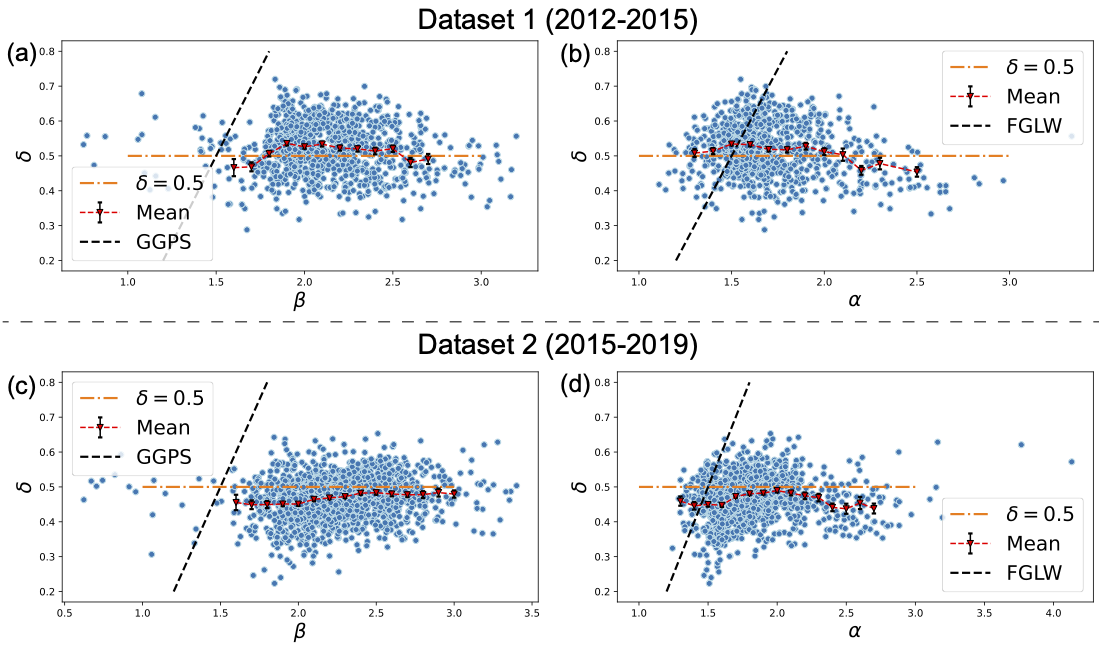}
			\caption{
				Scatterplots for the two-fold cross-validation regarding the GGPS and FGLW predictions.
				Fig.~(a, b) for Dataset 1 (from 2012 to 2015) and Fig.~(c, d) for Dataset 2 (from 2015 to 2019). 
			}
			\label{fig:FGLWGGPS-20122015-20152019-Delta}
			\end{figure*}
			Figure~\ref{fig:FGLWGGPS-20122015-20152019-Delta} shows the scatterplots to test the GGPS and FGLW predictions by using Dataset 1 (from 2012 to 2015, Fig.~\ref{fig:FGLWGGPS-20122015-20152019-Delta}(a, b)) and Dataset 2 (from 2015 to 2018, Fig.~\ref{fig:FGLWGGPS-20122015-20152019-Delta}(c, d)). These results are essentially the same as the main results in Fig.~\ref{fig:FGLW-GGPS-Delta} and play the role of the two-fold cross validation. Thus, the robustness of our analysis was affirmatively checked.

	\section{Statistical properties of our statistical methods in finite sample size}~\label{app:sec:simulation}
		In this Appendix, the statistical errors in estimating the power-law exponent $\delta_s$  are examined by considering a simple numerical simulation. Particularly, we study two issues: (i) {\it unbiasedness} of our estimators and (ii) the standard deviation of $\delta_s$ and $\chi_s$ due to finite sample-size effect. 

		Unbiasedness is one of the main desirable characters of estimators: for an estimator $\hat{\theta}$ with a true value $\theta$ and a sample size $N$, the estimator is called unbiased if its average is equal to the true value even for a finite sample size:
		\begin{equation}
			\la \hat{\theta}\ra = \theta \mbox{ for }N<\infty. 
		\end{equation}
		While it is feasible to construct unbiased estimators for simple setups (e.g., independently and identically distributed (IID) random variables), constructing exactly unbiased estimators for more complex setups, such as high-dimensional time series models, is challenging. Therefore, we numerically evaluate whether our estimator approximately satisfies unbiasedness by assuming a generative stochastic process for the price impacts.

		\subsection{Motivation to construct a plausible statistical model}~\label{app:sec:simulation-motivation}
			\begin{figure}
				\centering
				\includegraphics[width=70mm]{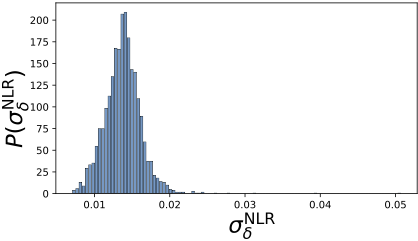}
				\caption{
					The histogram of the SEMs of individual stocks estimated by the standard nonlinear regression~\cite{NLRegression2007}. The typical size of the SEM is given by $\sigma_{\delta}^{\rm NLR}\approx 0.014$, which is much smaller than our final errorbar $\dla \overline{\sigma_{\delta}} \dra \approx 0.06$. This discrepancy is reasonable because the standard nonlinear regression assumes the IID observations of $(\mcQ,\mcI)$, which is incorrect for our time-series data.
				}
				\label{fig:Errorbars_NonlinearRegression}
			\end{figure}		 
			\begin{figure*}
				\centering
				\includegraphics[width=160mm]{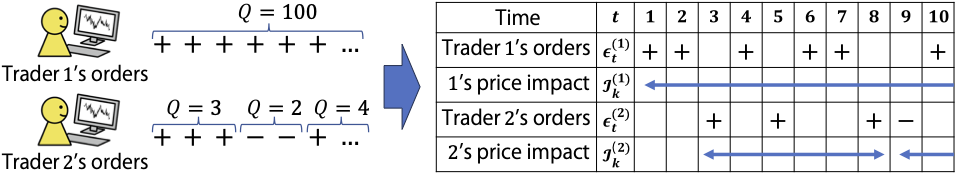}
				\caption{
					Correlations between price-impact observations $(\mcQ,\mcI)$. Price impacts between Traders 1 and 2 have correlation because of their temporal overlap.
				}
				\label{fig:SerialCorrealtion}
			\end{figure*}
			For reliable statistical estimation, it is necessary to construct a plausible statistical model because our dataset is a time-series dataset, where each observation is correlated and the typical statistical assumption of IID observations is violated\footnote{
				For example, the nonlinear regression algorithm~\cite{NLRegression2007} is often used for estimating power-law exponents by assuming the IID observations. Figure~\ref{fig:Errorbars_NonlinearRegression} illustrates the errorbar histogram estimated by the simple nonlinear least squares for the TSE dataset. The typical size of the individual errorbars was $\sigma_{\delta}^{\rm NLR}\approx 0.014$ among all stocks, which is much smaller than our final estimation $\dla \overline{\sigma_\delta} \dra \approx 0.06$. We did not use this standard nonlinear regression due to the non-IID nature of our dataset.
			}. In time-series analysis, such a non-IID problem is solved by developing statistical models generally. In econometrics, for example, the autoregressive integrated moving average (ARIMA) model has been developed in addressing such non-IID observation data, where the SEMs of the model parameters can be appropriately estimated\footnote{
				Remarkably, the OLS errorbars are underestimated than the ARIMA errorbars because the IID observations are wrongly assumed in the OLS model. Intuitively, the non-IID nature of the observations implies that the effective number of total observations is smaller than the superficial number of observations. This phenomenon is parallel to our results: the errorbar based on the standard nonlinear regression $\sigma_{\delta}^{\rm NLR}$---where the IID observations are assumed---is much smaller than that based on our final statistical model $\dla \overline{\sigma_\delta} \dra$---where the non-IID nature of our observations is taken into account: $\sigma_{\delta}^{\rm NLR}\ll \dla \overline{\sigma_\delta} \dra$.
			} even if the observations are non-IID. 
			
			More specifically, there are temporal overlaps between multiple metaorders in our dataset (see Fig.~\ref{fig:SerialCorrealtion}). Suppose that Trader 1 starts a buy metaorder from $t=1$ to $t=300$ with the volume $\mcQ_1=100$ and the resulting price impact $\mcI^{(1)}$. Also, suppose that Trader 2 starts a buy order from $t=3$ to $t=8$ with $\mcQ_2=3$ and the price imapct $\mcI^{(2)}$. Obviously, the price impacts $\mcI^{(1)}$ and $\mcI^{(2)}$ are not independent because of their temporal overlap. This is a candidate reason why $\sigma_{\delta}^{\rm NLR}$ might underestimate the actual errorbar. We thus conclude that proposing an appropriate statistical model is necessary to estimate the $\delta$'s errorbars based on our time-series data, similarly to econometrics.

		\subsection{Numerical models}~\label{app:sec:simulation-model}
			As a plausible statistical model, let us consider a generative stochastic model of the price impact with an exact half exponent $\delta = 1/2$, based on our real dataset. The idea of our model is straightforward: (i)~All metaorder execution schedules (in physical time) are identical to those in our real dataset. (ii)~However, the signs of the metaorders are randomly shuffled using our real dataset. (iii)~Furthermore, the price impact of each metaorder is modeled as a Gaussian random variable, following the square-root law on average with the exact half exponent $\delta = 1/2$. The randomness in our model arises from the random shuffling in process~(ii) and the Gaussian random variables in process~(iii). Since our model exhibits the square-root law price impact with an exact half exponent $\delta = 1/2$, it is feasible to evaluate the estimation error of $\delta$ from our real data attributable to the finite sample-size effect, by assuming the exact square-root law. In the following, we fix a stock $s$ and an operating day $j$, but omit the stock symbol $s$ and the operating day $j$ from the subscripts for simplicity.

\begin{figure*}
				\centering
				\includegraphics[width=160mm]{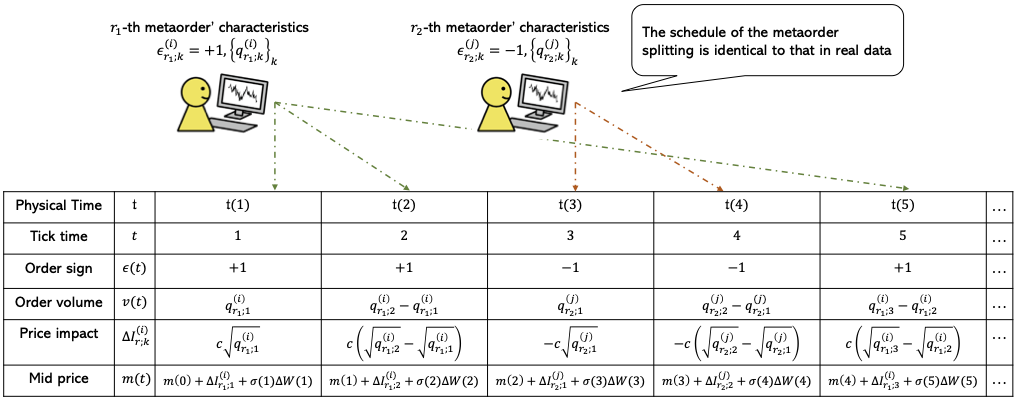}
				\caption{
					Schematic of our numerical simulation. The schedule of metaorder splitting is identical to that in our real dataset, but the metaorder signs are randomly shuffled. The price impact of each metaorder is designed to follow the exact square-root law on average. The model parameters, $c$ and $\sigma_{\Day}$, were calibrated using the real dataset. With this model, we repeatedly generated price time series to statistically evaluate the measurement errors attributable to the finite sample-size effect.}
				\label{fig:SimulationSchematic}
			\end{figure*}

			Let $t$ be the tick time, which increments by one with each metaorder submission. For a tick time $t$, the corresponding physical time is denoted by $\mathfrak{t}(t)$ [sec], which is identical to that in our real dataset. At the tick time $t$, we assume that trader $i$ submits his/her $k$-th child order belonging to the $r$-th metaorder. By the metaorder submission at the tick tim $t$, the market midprice after the transaction $m(t)$ obeys 
			\begin{equation}
				m(t) = m(t-1) + \Delta I_{r;k}^{(i)} + \sigma(t) \Delta W(t), \>\>\> \sigma^2(t) := \frac{\mathfrak{t}(t)-\mathfrak{t}(t-1)}{18,000 \mbox{ [sec]}},
			\end{equation}  
			where the price impact is designed to follow the exact square-root law on average, such that 
			\begin{equation}
				\Delta I_{r;k}^{(i)} = c\sigma_{\Day}\eps_{r;k}^{(i)}\left\{\left(q_{r;k}^{(i)}\right)^{1/2}-\left(q_{r;k-1}^{(i)}\right)^{1/2}\right\} \mbox{ for $k\geq 1$}, \>\>\> 
				q_{r;0}^{(i)} := 0,
			\end{equation}
			where $q_{r;k}^{(i)}$ is the cumulative volume up to the $k$-th child order belonging to the $r$-th metaorder for trader $i$ (i.e., $0=q_{r;0}^{(i)}<q_{r;1}^{(i)}< \dots \leq Q_{r}^{(i)}$, see Fig.~\ref{fig:TotalImpact-PathImpact-Schematic} for a schematic), and $\Delta W(t)$ is the standard white Gaussian noise term. Here, $c$ and $\sigma_{\Day}$ were set to be identical to those in the real dataset: $c$ was measured using the relative least squares for the price impact formula $\mcI(Q)=cQ^{1/2}$, and $\sigma_{\Day}$ represents the intraday volatility for the stock on the given operating day. Also, $\eps_{r;k}^{(i)}$ is the metaorder sign that is randomly shuffled across all traders, all metaorders, and all operating day. The white Gaussian noise term is added to maintain consistency with the standard nature of random walks of the price except for the effect of square-root price impacts. See Fig.~\ref{fig:SimulationSchematic} for the simulation schematic.

		\subsection{Numerical evaluation of the finite sample-size effect}~\label{app:sec:simulation-results}
			We can repeatedly generate price time series through numerical simulations of this model, in which the price impact follows the exact square-root law on average. We then measure $\delta$ in the same manner as for our real dataset and estimate the statistical error of $\delta$ attributable to the finite sample-size effect.

			Based on the simulation model, we repeatedly generated time series for $N_{\rm MC}$ iterations with $N_{\rm MC}=100$, 
			measured $\delta_{s;l}$ for all stocks on the $l$-th Monte Carlo iteration, and calculated their averages across stocks. For the $l$-th Monte Carlo trial, the average value of $\delta_{s;l}$ across stocks is denoted by $\overline{\delta_{s;l}}$. By aggregating all the Monte Carlo trials, the final averages across all stocks and trials are represented by $\dla \overline{\delta_s} \dra$. In other words, for any stochastic variable $A_{s;l}$, we define 
			\begin{equation}
				\overline{A_{s;l}} := \frac{1}{|\bm{\Omega}_{\rm S}|} \sum_{s \in \bm{\Omega}_{\rm S}} A_{s;l}, \>\>\>
				\dla \overline{A_s} \dra:= \frac{1}{N_{\rm MC}}\sum_{l=1}^{N_{\rm MC}} \overline{A_{s;l}}.
			\end{equation}
			
			We numerically obtained the summary statistics for the price impact, 
			\begin{equation}
				\dla \overline{\delta_s} \dra = 0.489 \pm 0.0013, \>\>\>
				\dla \overline{\sigma_{\delta}} \dra = 0.063,
			\end{equation}
			where $\dla \overline{\sigma_{\delta}} \dra$ is the standard deviation from the true value $1/2$ across all stocks and all trials
			\begin{equation}
				\dla \overline{\sigma_{\delta}} \dra
				:= \sqrt{\dla \overline{(\delta_s-1/2)^2}\dra}
				= \sqrt{\frac{1}{N_{\rm MC}}\sum_{l=1}^{N_{\rm MC}} \overline{\left(\delta_{s;l} - \frac{1}{2}\right)^2}} = \sqrt{\frac{1}{N_{\rm MC}}\sum_{l=1}^{N_{\rm MC}} \frac{1}{|\bm{\Omega}_{\rm S}|}\sum_{s\in \bm{\Omega}_{\rm S}}\left(\delta_{s;l} - \frac{1}{2}\right)^2}.
			\end{equation}
			Our numerical results imply that the average bias of our estimator $\delta_s$ is given by $|\dla \overline{\delta_s} \dra - 1/2| \approx 0.01$, which is very small. 

			In the same manner, we numerically evaluated the average and standard deviation for the exponent $\delta_s^{(i)}$ at the trader level, such that 
			\begin{align}
				\dla \overline{\delta_s^{(i)}} \dra = 0.521 \pm 0.0048, \>\>\>
				\dla \overline{\sigma_{\delta_s^{(i)}}} \dra = 0.169.
			\end{align}

			\subsection{Remark on the plausibility of our model}
			Whether the statistical model is plausible enough to capture reality is an important topic. In our recent preprint~\cite{NLProp2025}, we have exactly solved an essentially identical model with small adjustment and have shown that our statistical model is very plausible, consistent with various stylized facts (e.g., the diffusive nature of price dynamics, the inverse-cubic law, and the volatility clustering). Interested readers are referred to Ref.~\cite{NLProp2025} for more details. 
